\documentclass[a4paper,11pt]{article}
\pdfoutput=1
\usepackage{jheppub}

\usepackage{amssymb,amsmath}
\usepackage[normalem]{ulem}
\usepackage[utf8x]{inputenc}
\usepackage{slashed}
\usepackage{graphicx}
\usepackage{here}
\usepackage{color}
\usepackage{csquotes} 
\usepackage{comment}
\usepackage{mathrsfs}
\usepackage{float}
\usepackage{ascmac}
\usepackage{multirow}
\usepackage{longtable}

\usepackage[italicdiff]{physics}

\usepackage[hang,small,bf]{caption}
\usepackage[subrefformat=parens]{subcaption}
\makeatletter
\newcommand*\rel@kern[1]{\kern#1\dimexpr\macc@kerna}
\newcommand*\widebar[1]{%
  \begingroup
  \def\mathaccent##1##2{%
    \rel@kern{0.8}%
    \overline{\rel@kern{-0.8}\macc@nucleus\rel@kern{0.2}}%
    \rel@kern{-0.2}%
  }%
  \macc@depth\@ne
  \let\math@bgroup\@empty \let\math@egroup\macc@set@skewchar
  \mathsurround\z@ \frozen@everymath{\mathgroup\macc@group\relax}%
  \macc@set@skewchar\relax
  \let\mathaccentV\macc@nested@a
  \macc@nested@a\relax111{#1}%
  \endgroup
}
\makeatother

\numberwithin{equation}{section}

\preprint{
\begin{minipage}{5cm}
\small
\flushright
EPHOU-24-002\\
KEK-TH-2601
\\
KYUSHU-HET-281
\end{minipage}}

\title{Modular forms and hierarchical Yukawa couplings in 
heterotic Calabi-Yau compactifications}

\author{Keiya Ishiguro$^{1}$,} 
\author{Tatsuo Kobayashi$^{2}$,} 
\author{Satsuki Nishimura$^{3}$, and} 
\author{Hajime Otsuka$^{3}$} 
\affiliation{
$^1$Graduate University for Advanced Studies (Sokendai), 1-1 Oho, Tsukuba, Ibaraki 305-0801, Japan}
\affiliation{
$^2$Department of Physics, Hokkaido University, Sapporo 060-0810, Japan}
\affiliation{
$^3$Department of Physics, Kyushu University, 744 Motooka, Nishi-ku, Fukuoka 819-0395, Japan}
\emailAdd{ishigu@post.kek.jp}
\emailAdd{kobayashi@particle.sci.hokudai.ac.jp}
\emailAdd{nishimura.satsuki@phys.kyushu-u.ac.jp}
\emailAdd{otsuka.hajime@phys.kyushu-u.ac.jp}

\abstract{
We study the modular symmetry in heterotic string theory on Calabi-Yau threefolds. 
In particular, we examine whether moduli-dependent holomorphic Yukawa couplings are described by modular forms 
in the context of heterotic string theory with standard embedding. 
We find that $SL(2,\mathbb{Z})$ modular symmetry emerges in asymptotic regions of the Calabi-Yau moduli 
space. 
The instanton-corrected holomorphic Yukawa couplings are then given by modular forms under $SL(2,\mathbb{Z})$ or 
its congruence subgroups such as $\Gamma_0(3)$ and $\Gamma_0(4)$. 
In addition to the modular symmetry, it turns out that another coupling selection rule controls the structure of holomorphic Yukawa couplings. 
Furthermore, the coexistence of both the positive and negative modular weights for matter fields 
leads to a hierarchical structure of matter field K\"ahler metric. 
Thus, these holomorphic modular forms and the matter field K\"ahler metric play an important role in realizing a hierarchical 
structure of physical Yukawa couplings. 
}

\makeatletter
\gdef\@fpheader{}
\makeatother

\begin{document}

\maketitle

\section{Introduction}

The modular symmetry plays a crucial role in string theory. 
It was known that the one-loop closed bosonic string amplitude defined over the quotient of 
the upper half-plane of $SL(2,\mathbb{Z})$ is free of ultra-violet divergence. 
This discussion can be extended to higher-loop orders.\footnote{For more details, see, e.g., Ref.~\cite{Kaku:1999yd}.} For genus $h$ Riemann surfaces, the $SL(2,\mathbb{Z})$ 
modular group is replaced by the $Sp(2h,\mathbb{Z})$ modular group. 
The modular symmetry in the string worldsheet can also be seen in the target space of string theory. 

In toroidal compactifications, the modular symmetry can be regarded as a geometric symmetry of torus, 
which puts strong constraints on the four-dimensional (4D) low-energy effective field theory. 
Indeed, the modulus $\tau$ parametrizing the $SL(2,\mathbb{Z})$ modular symmetry appears in couplings of matter fields, 
as known in heterotic orbifold models \cite{Ferrara:1989qb,Lerche:1989cs,Lauer:1989ax,Lauer:1990tm} and magnetized D-brane models \cite{Kobayashi:2018rad,Kobayashi:2018bff,Ohki:2020bpo,Kikuchi:2020frp,Kikuchi:2020nxn,
Kikuchi:2021ogn,Almumin:2021fbk,Kikuchi:2023awe}. 
In these models, chiral zero-modes have certain representations of the congruence subgroup of $SL(2,\mathbb{Z})$ with 
a modular weight, and these holomorphic Yukawa couplings are described by a certain modular form under the congruence 
subgroup. It is remarkable that the inhomogeneous modular group $\bar{\Gamma} = PSL(2,\mathbb{Z})$ has several 
finite subgroups such as $S_3, A_4, S_4, A_5$ \cite{deAdelhartToorop:2011re}. 
These modular groups will be important to understand the flavor structure of 
quarks and leptons \cite{Altarelli:2010gt,Ishimori:2010au,Kobayashi:2022moq,Hernandez:2012ra,King:2013eh}.

In Calabi-Yau (CY) compactifications as an extension of toroidal backgrounds\footnote{See, Ref.~\cite{Kikuchi:2022txy}, for a construction of 
modular symmetric models from higher-dimensional theory on general grounds.}, 
the modular symmetry is enhanced to $Sp(2h+2,\mathbb{Z})$ with $h$ being the number of moduli fields~\cite{Strominger:1990pd,Candelas:1990pi}. 
Such a symplectic modular symmetry will constrain the 4D low-energy effective action~\cite{Ishiguro:2020nuf,Ishiguro:2021ccl}. 
It was pointed out in Ref.~\cite{Ishiguro:2020nuf} that the flavor symmetry of matter fields can be embedded into the 
symplectic modular symmetry in the context of heterotic string theory with standard embedding. 
However, it is still unclear whether holomorphic Yukawa couplings are described by modular forms. 
The derivation of holomorphic Yukawa couplings is also important to understand the hierarchical structure of 
quark and lepton masses.\footnote{See for the derivation of holomorphic Yukawa couplings 
for heterotic line bundle models \cite{Blesneag:2015pvz}.} 
The purpose of this paper is to derive explicit forms of holomorphic Yukawa couplings in heterotic 
string theory on concrete CY threefolds. 
We find that the $SL(2,\mathbb{Z})$ modular symmetry emerges in asymptotic regions of the complex structure and  
K\"ahler moduli spaces. The instanton-corrected holomorphic Yukawa couplings are then described by modular forms 
of $SL(2,\mathbb{Z})$ or its congruence subgroups such as $\Gamma_0(3)$ and $\Gamma_0(4)$. 
Furthermore, it turns out that these holomorphic modular forms as well as K\"ahler metric of matter fields can 
lead to the hierarchical structure of physical Yukawa couplings.

This paper is organized as follows. 
In Sec. \ref{sec:review}, we review the 4D low-energy effective field theory of heterotic string theory with standard 
embedding, focusing on the role of modular symmetry. 
In Sec. \ref{sec:modular_forms}, we study instanton-corrected holomorphic Yukawa couplings on several CY threefolds. 
The structure of instanton numbers determines the modular form of holomorphic Yukawa couplings such as modular forms of $SL(2,\mathbb{Z})$ in Sec. \ref{sec:sl2}, 
$\Gamma_0(3)$ in Sec. \ref{sec:gamma03} and $\Gamma_0(4)$ in Sec. \ref{sec:gamma04}. 
As an application of modular forms, we investigate whether one can realize the hierarchical structure of physical Yukawa couplings 
in Sec. \ref{sec:hierarchy}. 
Sec. \ref{sec:con} is devoted to conclusions. 
In Appendix \ref{sec:app}, we summarize several holomorphic modular forms under the $\Gamma_0(N)$ modular group.

\section{Modular invariant effective action}
\label{sec:review}

In this section, we briefly review the 4D low-energy effective field theory action derived from  $E_8\times E_8^\prime$ heterotic string theory on smooth Calabi-Yau threefolds with standard embedding.\footnote{The discussion is the same for $SO(32)$ heterotic string theory.}
In particular, we emphasize the role of symplectic modular symmetry in the effective action of matter fields which are in one-to-one correspondence with the moduli fields. 

\paragraph{K\"ahler potential}

It is known that the low-energy group is described by $E_6\times E_8^\prime$ due to the identification of background $SU(3) \subset E_8$ 
gauge field with the spin connection of CY threefolds. 
There are $h^{1,1}$ number of chiral superfields $A^a$ in the ${\bf 27}$ representation of $E_6$ and $h^{2,1}$ number of chiral superfields $A^i$ in the $\widebar{{\bf 27}}$ representation of $E_6$. 
Here, $h^{1,1}$ and $h^{2,1}$ denote the hodge numbers of CY threefolds. 
Thanks to the one-to-one correspondence between moduli and chiral matter fields, the matter field K\"ahler metrics are written by  the following form~\cite{Dixon:1989fj}\footnote{Note that the overall factor $e^{\pm\frac{1}{3}(K_{\rm cs}-K_{\rm ks})}$ 
is different in toroidal orbifolds due to the enlarged symmetry \cite{Dixon:1989fj}.}:
\begin{align}
    K^{({\bf 27})}_{a\widebar{b}} &= e^{\frac{1}{3}(K_{\rm cs}-K_{\rm ks})}(K_{\rm ks})_{a\widebar{b}},
    \nonumber\\
   K^{(\widebar{{\bf 27}})}_{i\widebar{j}} &= e^{-\frac{1}{3}(K_{\rm cs}-K_{\rm ks})}(K_{\rm cs})_{i\widebar{j}},
   \label{eq:KahlerMetric}
\end{align}
where $(K_{\rm ks})_{a\widebar{b}}=\partial_a\partial_{\bar{b}}K_{\rm ks}$ and $(K_{\rm cs})_{i\widebar{j}}=\partial_i\partial_{\bar{j}}K_{\rm cs}$ are the metric of the K\"ahler moduli and the complex structure moduli, respectively. 
To see an explicit form of the matter field K\"ahler metric, let us focus on the K\"ahler moduli sector for concreteness. 
The K\"ahler potential of the K\"ahler moduli is well controlled in the large volume regime ${\cal V}\gg l_s^6$
\begin{align}
    K_{\rm ks} 
    = -\ln {\cal V} = - \ln \biggl[i \frac{\kappa_{abc}}{6} (t^a - \widebar{t}^a)(t^b - \widebar{t}^b)(t^c - \widebar{t}^c)\biggl].
\end{align}
Here, $\kappa_{abc}$ denotes a triple intersection number of CY threefolds. 
It results in the K\"ahler metric of ${\bf 27}$ matters:
\begin{align}
K^{({\bf 27})}_{a\widebar{b}} &= e^{\frac{1}{3}K_{\rm cs}}{\cal V}^{1/3}(K_{\rm ks})_{a\widebar{b}}
\nonumber\\
& =e^{\frac{K_{\rm cs}}{3}}\biggl[\frac{1}{{\cal V}^{5/3}} \left( \frac{\kappa_{ade}}{2} (t^d - \widebar{t}^d)(t^e - \widebar{t}^e)\right)\left( \frac{\kappa_{bfg}}{2} (t^f - \widebar{t}^f)(t^g - \widebar{t}^g)\right)
+ \frac{i}{{\cal V}^{2/3}}\kappa_{abh}(t^h - \widebar{t}^h)
\biggl],
\label{eq:K27metric}
\end{align}
where we use
\begin{align}
    (K_{\rm ks})_{a\widebar{b}} &= \frac{1}{{\cal V}}\left( \frac{\partial_a {\cal V} \partial_{\bar b} {\cal V}}{{\cal V}} -\partial_a \partial_{\bar b}{\cal V} \right)
\nonumber\\
&    = \frac{1}{{\cal V}^2} \left( \frac{\kappa_{ade}}{2} (t^d - \widebar{t}^d)(t^e - \widebar{t}^e)\right)\left( \frac{\kappa_{bfg}}{2} (t^f - \widebar{t}^f)(t^g - \widebar{t}^g)\right)
+ \frac{i}{{\cal V}}\kappa_{abh}(t^h - \widebar{t}^h).
\end{align}
The K\"ahler metric of $\widebar{{\bf 27}}$ matters is written by the same expression, 
where the large volume regime corresponds to the large complex structure moduli. 
If we restrict K\"ahler moduli to the universal value, that is, $t:= t^i$ for all $i$, the matter field K\"ahler metric is proportional to 
\begin{align}
K^{({\bf 27})}_{a\widebar{b}} 
\propto  e^{\frac{1}{3}K_{\rm cs}} \frac{1}{i(\Bar{t}-t)}.
\end{align}
In the following, we focus on the effective action of ${\bf 27}$ matter fields, but a similar analysis can be done for the $\widebar {{\bf 27}}$ matters.

\paragraph{Yukawa couplings}

The 3-point coupling of moduli fields is given by the third derivative of the prepotential with respect to corresponding moduli fields. 
In the large volume regime, this 3-point coupling is nothing but the triple intersection numbers $\kappa_{abc}$ classically.
However, it will be corrected by instanton contributions. 
It is known that instanton-corrected Yukawa couplings ere obtained in the following form \cite{Strominger:1985ks,Hosono:1994ax}
\begin{align}
    y_{abc} &= \kappa_{abc} + \sum_{d_1, d_2,...,d_m=0}^\infty \frac{(d_a d_b d_c) n_{d_1,d_2,...,d_m}}{1- \Pi_{l=1}^m q_l^{d_l}} \Pi_{l=1}^m q_l^{d_l},
\end{align}
with $q_l = e^{2\pi i t_l}$, where the instanton numbers $n_{d_1,d_2,...,d_m}$ are determined once the background CY geometry is fixed. 
The ${\bf 27}$ matter superpotential is then written by 
\begin{align}
    W = y_{abc}A^a A^b A^c.
\end{align}

\paragraph{Modular symmetry}

In the large volume regime, the effective action is controlled by a certain subgroup of symplectic modular group $Sp(2h^{1,1}+2, \mathbb{Z})$, which depends on the structure of the intersection numbers $\kappa_{abc}$. The full modular group can be described in projective coordinates $Y^A$ with $A=1,2,...,h^{1,1}+1$. Here, the K\"ahler moduli correspond to
\begin{align}
t^a = \frac{Y^a}{Y^0},
\end{align}
with $a=1,2,...,h^{1,1}$. 
As discussed in Ref. \cite{Ishiguro:2021ccl}, the symplectic modular transformations of matter fields and their K\"ahler metric
\begin{align}
    A^a &\rightarrow \Tilde{A}^a=(\Tilde{Y}^0)^{-1/3} \frac{\partial \Tilde{t}^a}{\partial t^b}A^b,
    \nonumber\\
    K^{({\bf 27})}_{a\Bar{b}} &\rightarrow  \tilde{K}^{({\bf 27})}_{a\Bar{b}} = |\Tilde{Y}^0|^{2/3} \frac{\partial t^c}{\partial \Tilde{t}^a}  \frac{\partial \bar{t}^{\bar{d}}}{\partial \Tilde{\bar{t}}^{\bar{b}}} K^{({\bf 27})}_{c\Bar{d}},
\end{align}
are consistent with the modular symmetry of the K\"ahler moduli: $t^a\rightarrow \tilde{t}^a$. 
Furthermore, Yukawa couplings are expected to be transformed as
\begin{align}
    y_{abc} \rightarrow \tilde{y}_{abc} = \tilde{Y}^0 \frac{\partial t^d}{\partial \Tilde{t}^a} \frac{\partial t^e}{\partial \Tilde{t}^b} \frac{\partial t^f}{\partial \Tilde{t}^c} y_{def}.  
\end{align}
Thus, the symplectic modular symmetry will determine the flavor symmetry of matter fields.

Note that instanton effects may further break the remained modular group into a smaller one. 
The tree-level Yukawa couplings $\kappa_{abc}$ are moduli-independent constants and correspond to trivial modular forms of weight zero. 
On the other hand, moduli-dependent holomorphic Yukawa coupling terms induced by instanton effects can be non-trivial modular forms.
That is interesting from both theoretical and phenomenological viewpoints.
In the following section, we study explicitly several CY threefolds in order to
reveal unbroken modular symmetries and concrete forms of 
non-trivial modular forms.

\section{Holomorphic Yukawa couplings}
\label{sec:modular_forms}

In this section, we study a modular invariance of the effective action for matter fields. 
In Sec. \ref{sec:sl2}, we first derive a $SL(2,\mathbb{Z})$ modular form for holomorphic 
Yukawa couplings of matter fields, which are realized in specific CY threefolds with two moduli in Sec. \ref{sec:sl2_two} and 
three moduli in Sec. \ref{sec:sl2_three}. 
In both examples, holomorphic Yukawa couplings are described by an Eisenstein series with weight 4 which transforms as a singlet representation under $SL(2,\mathbb{Z})$. 
In Sec. \ref{sec:gamma03}, we next show that the holomorphic Yukawa couplings are described by a modular form under $\Gamma_0(3)$ as a subgroup of $SL(2,\mathbb{Z})$, which are realized in specific CY threefolds with two moduli in Sec. \ref{sec:gamma03_two} and three moduli in Sec. \ref{sec:gamma03_three}.
Finally, we present the holomorphic Yukawa couplings controlled by $\Gamma_0(4)$ modular group in Sec. \ref{sec:gamma04}.  

\subsection{$SL(2,\mathbb{Z})$ modular forms}
\label{sec:sl2}

In this section, we focus on specific CY threefolds with two moduli in Sec. \ref{sec:sl2_two} and three moduli 
in Sec. \ref{sec:sl2_three}, leading to a $SL(2,\mathbb{Z})$ modular form for holomorphic Yukawa couplings.

\subsubsection{Two moduli}
\label{sec:sl2_two}

We first study a CY threefold defined by the degree 18 hypersurface in weighted projective space $\mathbb{CP}^{1,1,1,6,9}[18]$, 
where the number of K\"ahler moduli is $h^{1,1}=2$. 
We revisit the work of Ref.~\cite{Candelas:1994hw} from the viewpoint of effective action of matter fields. 
By solving the corresponding Picard-Fuchs equation, the prepotential for two K\"ahler moduli was known to be
\begin{align}
    {\cal F} &= -\frac{1}{6}\left(9 t_1^3 + 9t_1^2 t_2 +3 t_1 t_2^2 \right),
\end{align}
where we omit the higher-order corrections including instanton effects, and the triple intersection numbers are given by
\begin{align}
    \kappa_{111} = 9,\quad
    \kappa_{112}= 3,\quad
    \kappa_{122}=1,
\end{align}
and otherwise 0. 
Instanton-corrected holomorphic Yukawa couplings are of the form:
\begin{align}
    y_{abc} &= \kappa_{abc} + \sum_{d_1, d_2=0}^\infty \frac{c_{abc}(d_1,d_2)n_{d_1,d_2}q_1^{d_1} q_2^{d_2}}{1-q_1^{d_1} q_2^{d_2}},
\end{align}
with
\begin{align}
    c_{111}=(d_1)^3,\quad
    c_{112}=(d_1)^2 d_2,\quad    
    c_{122}=d_1(d_2)^2,\quad        
    c_{112}=(d_2)^3.    
\end{align}
Here, the instanton numbers $n_{d_1,d_2}$ are given in Table \ref{tab:instantons_1}.
\begin{table}[H]
    \centering
    \begin{tabular}{c|c|c|c|c|}
       $d_1$ $\setminus$ $d_2$  &  0  & 1 & 2 & 3 \\ \hline
       0  & & 3 & -6 & 27  \\
       1  & 540 & -1080 & 2700 & -17280  \\       
       2  & 540 & 143370 & -574560 & 5051970  \\       
       3  & 540 & 204071184 & 74810520 & -913383000 \\ \hline      
    \end{tabular}
    \caption{Instanton numbers up to $d_1, d_2 \leq 3$.}
    \label{tab:instantons_1}
\end{table}
Remarkably, instanton numbers $n_{d_1,0}$ are the same number 540 for all $d_1$. 
When we restrict ourselves to the regime $q_2 \ll q_1$ with $q_1=e^{2\pi it_1}$ and $q_2=e^{2\pi it_2}$ such that the $t_2$-dependent instanton corrections are suppressed, only the sector with $d_2=0$ contributes to the holomorphic Yukawa couplings. 
In this regime, holomorphic Yukawa couplings are described by an Eisenstein series with weight 4: 
\begin{align}
    y_{111} &= 9 + 540 \sum_{k=0}^\infty \frac{k^3 q_1^k}{1-q_1^k} = \frac{27}{4} + \frac{9}{4}E_4(t_1),\nonumber\\
    y_{112} &= 3,\nonumber\\
    y_{122} &= 1,\nonumber\\
    y_{222} &= 0,
\end{align}
where
\begin{align}
    E_4(t) = 1 + 240\sum_{k=0}^\infty \frac{k^3 q^k}{1-q^k}.
\end{align}
It indicates an existence of $SL(2,\mathbb{Z})$ modular symmetry, as confirmed at the level of period integrals \cite{Candelas:1994hw}. 
When we move to the following basis:
\begin{align}
    t :=t^1,\qquad
    s := \frac{3}{2}t^1 + t^2,
\end{align}
the $SL(2,\mathbb{Z})$ symmetry can appear manifest in the regime $q_s \ll q_t$ with $q_t :=q_1$ and $q_s := e^{2\pi i s}=q_1^{3/2}q_2$. 
In this basis, the prepotential is rewritten in terms of the redefined moduli as
\begin{align}
    {\cal F} &= -\frac{1}{6}\left(\frac{9}{4} t^3 + 3ts^2 \right).
\end{align}

To see the modular invariance of the effective action, let us consider the modular weights of matter fields, 
which can be read off from the moduli K\"ahler potential:
\begin{align}
    K_{\rm ks} &= - \ln \biggl[i \left(\frac{3}{8} (t - \widebar{t})^3 +\frac{1}{2}(t - \widebar{t})(s- \widebar{s})^2\right)\biggl].
\end{align}
Since the corresponding moduli K\"ahler metric is given in the regime ${\rm Im}(t) \ll {\rm Im}(s)$:
\begin{align}
    (K_{\rm ks})_{t\Bar{t}} \simeq - \frac{1}{(t-\widebar{t})^2},
    \qquad
    (K_{\rm ks})_{s\Bar{s}} \simeq - \frac{2}{(s-\widebar{s})^2},
    \qquad
    (K_{\rm ks})_{t\Bar{s}} = K_{s\Bar{t}} \simeq 0,
\end{align}
one can determine the matter modular weights from $t$-dependent part of the moduli K\"ahler metric by using Eq. (\ref{eq:KahlerMetric}),  
\begin{align}
    K^{({\bf 27})}_{t\widebar{t}} &\sim e^{\frac{1}{3}(-K_{\rm ks})}(K_{\rm ks})_{t\widebar{t}} \simeq \frac{1}{(t-\widebar{t})^{5/3}},
    \nonumber\\
    K^{({\bf 27})}_{s\widebar{s}} &\sim e^{\frac{1}{3}(-K_{\rm ks})}(K_{\rm ks})_{s\widebar{s}} \simeq (t-\widebar{t})^{1/3}.
\end{align}
It leads to the modular weights $-\frac{5}{3}$ for $A^{({\bf 27})}_t$ and $\frac{1}{3}$ for $A^{({\bf 27})}_s$. 
Then, the modular weights of holomorphic Yukawa couplings are determined, and their explicit forms are rewritten as
\begin{align}
    y_{ttt} &= \frac{9}{4}E_4(t)\qquad ({\rm weight}\,4), \nonumber\\
    y_{tts} &= 0\hspace{48pt} ({\rm weight}\,2),\nonumber\\
    y_{tss} &= 1\hspace{48pt} ({\rm weight}\,0),\nonumber\\
    y_{sss} &= 0\hspace{48pt} ({\rm weight}\,-2).
\end{align}
In this basis, they are described by modular forms of weights 4, 2, 0, -2, respectively.
Since there are no modular forms of weights 2 and -2 under $SL(2,\mathbb{Z})$, they are forced to be zero.
By definition, the modular form of weight 0 is just a modulus-independent constant. 
Thus, this result is consistent with the modular invariance of 4D effective action. 
Indeed, under the $SL(2,\mathbb{Z})_t$ modular transformation:
\begin{align}
    t \rightarrow \frac{at +b}{c t +d}
\end{align}
with $ad-bc=1$, the K\"ahler potential and superpotential 
\begin{align}
    K &= -\ln \biggl[\frac{i}{2}(t - \widebar{t})(s- \widebar{s})^2\biggl] +  K^{({\bf 27})}_{t\widebar{t}} |A^{({\bf 27})}_t|^2 
    + K^{({\bf 27})}_{s\widebar{s}} |A^{({\bf 27})}_s|^2,
    \nonumber\\
    W &= y_{ttt}A^{({\bf 27})}_tA^{({\bf 27})}_tA^{({\bf 27})}_t + y_{tss}A^{({\bf 27})}_t A^{({\bf 27})}_s A^{({\bf 27})}_s
\end{align}
transform as
\begin{align}
K \rightarrow K + \ln |ct +d|^2,\qquad
W \rightarrow (ct +d)^{-1}W,
\end{align}
and the K\"ahler invariant quantity $e^K |W|^2$ is modular invariant.

\subsubsection{Three moduli}
\label{sec:sl2_three}

We move to a CY threefold with a K3 fibration structure embedded in weighted projective spaces:
\begin{align}
\mathbb{CP} 
\begin{pmatrix}
    6 & 4 & 1 & 0 & 1 & 0
    \\ 
    6 & 4 & 0 & 1 & 0 & 1
\end{pmatrix}
\left[
\begin{array}{c}
     12  \\
     12
\end{array}
\right].
\end{align}
As studied in Ref.  \cite{Klemm:2004km}, the leading part of the prepotential for three K\"ahler moduli is described by
\begin{align}
    {\cal F} &= -\frac{4}{3}t_1^3 - t_1^2t_2 -t_1^2t_3  -t_1t_2t_3, 
\end{align}
where the triple intersection numbers are
\begin{align}
    \kappa_{111} = 8, \quad
    \kappa_{112}=2, \quad
    \kappa_{113}=2,\quad
    \kappa_{123}=1,
\end{align}
and otherwise 0. The holomorphic Yukawa couplings of matter fields are determined by the classical part and instanton 
contribution:
\begin{align}
    y_{abc} = \kappa_{abc} + \sum_{d_1, d_2, d_3=0}^\infty \frac{c_{abc}(d_1,d_2,d_3)n_{d_1, d_2, d_3}q_1^{d_1} q_2^{d_2}q_3^{d_3}}{1-q_1^{d_1} q_2^{d_2}q_3^{d_3}},
    \label{eq:Yukawa_3mod}
\end{align}
with
\begin{align}
    c_{abc}(d_1,d_2,d_3) = d_a d_b d_c,\quad
    q_1 = e^{2\pi i t_1},\quad 
    q_2 = e^{2\pi i t_2},\quad
    q_3 = e^{2\pi i t_3}.
\end{align}
Let us consider the regime $q_1, q_2 \ll q_3$; one can approximate the instanton expansion by focusing on $d_3=0$. 
The nonvanishing instanton numbers $n_{d_1, d_2, 0}$ are given in Table \ref{tab:instantons_2}.
\begin{table}[H]
    \centering
    \begin{tabular}{c|c|c|c|c|}
       $d_1$ $\setminus$ $d_2$  &  0  & 1 & 2 & 3 \\ \hline
       0  & & -2 &  &   \\
       1  & 480 & 480 &  &   \\       
       2  & 480 & 282888 & 480 &   \\       
       3  & 480 & 17058560 & 17058560 & 480 \\ \hline      
    \end{tabular}
    \caption{Nonvanishing instanton numbers up to $d_1, d_2 \leq 3$.}
    \label{tab:instantons_2}
\end{table}
As seen in the previous section, instanton numbers $n_{d_1,0}$ and $n_{d_1, d_1}$ are the same number 480 for all $d_1$. 
When we restrict ourselves to the regime $q_2 \ll q_1$ such that the $t_2$-dependent instanton corrections are suppressed, the sector with $d_2=0$ is relevant for holomorphic Yukawa couplings: 
\begin{align}
    y_{111} &= 8 + 480 \sum_{k=0}^\infty \frac{k^3 q^k_1}{1-q^k_1} = 6 + 2E_4(t_1),\nonumber\\
    y_{112} &= 2,\nonumber\\
    y_{113} &= 2,\nonumber\\
    y_{123} &= 1.
    \label{eq:hol_yukawa}
\end{align}
Since an Eisenstein function with weight 4 is a modular function under $SL(2,\mathbb{Z})$, 
it indicates the $SL(2,\mathbb{Z})$ modular symmetry for the effective action of matter fields. 
To check the modular invariance of the effective action, let us redefine the moduli as
\begin{align}
    t := t_1,
    \qquad
    u := t_1 + t_2,
    \qquad
    s :=t_1 + t_3.
    \label{eq:stu}
\end{align}
In this basis, the prepotential becomes
\begin{align}
    {\cal F} &= -\frac{4}{3}t_1^3 - t_1^2t_2 -t_1^2t_3  -t_1t_2t_3
    \nonumber\\
    &= -\frac{1}{6}\left(2t^3 + 6 tus\right).
\end{align}
The limit $q_{2}\rightarrow 0$ and $q_3 \rightarrow 0$ correspond to $q_u = e^{2\pi i u} = q_2 q_t \rightarrow 0$ and $q^s = e^{2\pi i t_3}=q_3 \rightarrow 0$, respectively. 
The moduli K\"ahler potential is also given by
\begin{align}
    K_{\rm ks} &= - \ln \biggl[ i \left(\frac{1}{3} (t - \widebar{t})^3 + (t - \widebar{t})(u- \widebar{u})(s- \widebar{s})\right)\biggl],
\end{align}
and the corresponding moduli K\"ahler metric in  the regime ${\rm Im}(t) \ll {\rm Im}(s),{\rm Im}(u)$:
\begin{align}
    (K_{\rm ks})_{t\Bar{t}} &\simeq - \frac{1}{(t-\widebar{t})^2},
    \qquad
    (K_{\rm ks})_{s\Bar{s}} \simeq - \frac{1}{(s-\widebar{s})^2},
    \nonumber\\
    (K_{\rm ks})_{u\Bar{u}} &\simeq - \frac{1}{(u-\widebar{u})^2},
\end{align}
and otherwise 0. 
By using Eq. (\ref{eq:KahlerMetric}), one can determine the matter modular weights from $t$-dependent part of the moduli K\"ahler metric: 
\begin{align}
    K^{({\bf 27})}_{t\widebar{t}} &\sim e^{\frac{1}{3}(-K_{\rm ks})}(K_{\rm ks})_{t\widebar{t}} \simeq \frac{1}{(t-\widebar{t})^{5/3}},
    \nonumber\\
    K^{({\bf 27})}_{u\widebar{u}} &\sim e^{\frac{1}{3}(-K_{\rm ks})}(K_{\rm ks})_{u\widebar{u}} \simeq (t-\widebar{t})^{1/3},
    \nonumber\\
    K^{({\bf 27})}_{s\widebar{s}} &\sim e^{\frac{1}{3}(-K_{\rm ks})}(K_{\rm ks})_{s\widebar{s}} \simeq (t-\widebar{t})^{1/3},
\end{align}
that is, modular weight $-\frac{5}{3}$ for $A^{({\bf 27})}_t$ and $\frac{1}{3}$ for $A^{({\bf 27})}_s$ and $A^{({\bf 27})}_u$. 
Then, the modular weights of holomorphic Yukawa couplings are determined, and their explicit forms are rewritten as
\begin{align}
    y_{ttt} &= 2 E_4(t)\qquad({\rm weight}\,4),\nonumber\\
    y_{stu}& =1\hspace{48pt}({\rm weight}\,0),
\end{align}
and otherwise 0. 
In this basis, they are described by modular forms of weights 4, 0, respectively. 
Thus, this result is consistent with the modular invariance of 4D effective action. 
Indeed, under the $SL(2,\mathbb{Z})_t$ modular transformation:
\begin{align}
    t \rightarrow \frac{at +b}{c t +d}
\end{align}
with $ad-bc=1$, the K\"ahler potential and superpotential 
\begin{align}
    K &= - \ln \biggl[i \left((t - \widebar{t})(u- \widebar{u})^2+(t - \widebar{t})(u- \widebar{u})(s- \widebar{s})\right)\biggl] +  \sum_{a,b} K^{({\bf 27})}_{a\widebar{b}} A^{({\bf 27})}_a \widebar{A^{({\bf 27})}_b},
    \nonumber\\
    W &= y_{ttt}A^{({\bf 27})}_tA^{({\bf 27})}_tA^{({\bf 27})}_t + y_{stu}A^{({\bf 27})}_s A^{({\bf 27})}_t A^{({\bf 27})}_u
\end{align}
transform as
\begin{align}
K \rightarrow K + \ln |ct +d|^2,\qquad
W \rightarrow (ct +d)^{-1}W,
\label{eq:sl2z_below}
\end{align}
and the K\"ahler invariant quantity $e^K |W|^2$ is modular invariant. 

Note that this effective theory also seems to have another coupling selection rule.
The matter fields $A^{({\bf 27})}_s$ and $A^{({\bf 27})}_u$ have the same modular weight.
The modular invariance allows $y_{sst}A^{({\bf 27})}_s A^{({\bf 27})}_s A^{({\bf 27})}_t$ and 
$y_{tuu}A^{({\bf 27})}_t A^{({\bf 27})}_u A^{({\bf 27})}_u$ couplings.
For example, if $A^{({\bf 27})}_s$ and $A^{({\bf 27})}_u$ are $S_4$ doublet and 
$A^{({\bf 27})}_t$ is a $S_4$ singlet,  $y_{sst}A^{({\bf 27})}_s A^{({\bf 27})}_s A^{({\bf 27})}_t$ and 
$y_{tuu}A^{({\bf 27})}_t A^{({\bf 27})}_u A^{({\bf 27})}_u$ couplings are forbidden, but 
$y_{ttt}A^{({\bf 27})}_tA^{({\bf 27})}_tA^{({\bf 27})}_t$ and  $y_{stu}A^{({\bf 27})}_s A^{({\bf 27})}_t A^{({\bf 27})}_u$ 
are allowed \cite{Kobayashi:2022moq}.
One can replace $S_4$ by $S_3$.
The $S_4$ group as well as the $S_3$ group is originated from $Sp(8,\mathbb{Z})$.

\subsection{$\Gamma_0(3)$ modular forms}
\label{sec:gamma03}

In this section, we focus on specific CY threefolds with two moduli in Sec. \ref{sec:gamma03_two} and three moduli 
in Sec. \ref{sec:gamma03_three}, leading to $\Gamma_0(3)$ modular form for holomorphic Yukawa couplings. 
The $\Gamma_0(N)$ modular forms are summarized in Appendix \ref{sec:app}.

\subsubsection{Two moduli}
\label{sec:gamma03_two}

We start with a CY threefold defined by a hypersurface in a product of two  projective spaces:
\begin{align}
\begin{matrix}
\mathbb{CP}^{2}\\
\mathbb{CP}^{2}\\
\end{matrix}
\begin{bmatrix}
3\\
3\\
\end{bmatrix}
,
\end{align} 
where the number of K\"ahler moduli is $h^{1,1}=2$. 
The moduli effective action was studied in Ref.~\cite{Hosono:1993qy} by using the mirror symmetry technique. 
Here, we study whether the effective action of matter fields enjoys the modular symmetry. 
By solving the corresponding Picard-Fuchs equation, the leading part of the prepotential for two K\"ahler moduli was known to be
\begin{align}
    {\cal F} &= -\frac{1}{6}\left(9t_1^2 t_2 +9 t_1 t_2^2 \right),
\end{align}
where the triple intersection numbers are given by
\begin{align}
    \kappa_{111} =\kappa_{222} = 0,\quad
    \kappa_{112}= \kappa_{122} =3.
\end{align}
Instanton-corrected holomorphic Yukawa couplings are of the form:
\begin{align}
    y_{abc} &= \kappa_{abc} + \sum_{d_1, d_2=0}^\infty \frac{c_{abc}(d_1,d_2)n_{d_1,d_2}q_1^{d_1} q_2^{d_2}}{1-q_1^{d_1} q_2^{d_2}},
\end{align}
with
\begin{align}
    c_{111}=(d_1)^3,\quad
    c_{112}=(d_1)^2 d_2,\quad    
    c_{122}=d_1(d_2)^2,\quad        
    c_{112}=(d_2)^3.    
\end{align}
Here, the instanton numbers $n_{d_1,d_2}$ are given in Table. \ref{tab:instantons_gamma03_1}.
\begin{table}[H]
    \centering
    \begin{tabular}{c|c|c|c|c|c|c|c|}
       $d_1$ $\setminus$ $d_2$  &  0  & 1 & 2 & 3 & 4 & 5 & 6 \\ \hline
       0  &  & 189 & 189 & 162 & 189 & 189 & 162   \\
       1  & 189 & 8262 & 142884 & 1492290 & 11375073 & 69962130 & \\       
       2  & 189 & 142884 & 13108392 & 516953097 & 12289326723 & & \\
       3  & 162 & 1492290 & 516953097 & 55962304650 & & & \\ 
       4  & 189 & 11375073 & 12289326723 & & & & \\
       5  & 189 & 69962130 & & & & & \\
       6  & 162 & & & & & & \\ 
       \hline
    \end{tabular}
    \caption{Instanton numbers up to bidegree $d_1+d_2 \leq 6$. Note that there exists the exchange symmetry $n_{d_1, d_2} = n_{d_2, d_1}$.}
    \label{tab:instantons_gamma03_1}
\end{table}
Remarkably, instanton numbers $n_{d_1,0}$ have an interesting structure such as 
$n_{d_1,0} = 162$ for $d_1 = 3 \mathbb{Z}_+$ and otherwise 189. 
Thus, when we restrict ourselves to the regime $q_2 \ll q_1$ such that the $t_2$-dependent instanton corrections are suppressed, only $t_1$-dependent instantons contribute to the holomorphic Yukawa couplings. 
In particular, one of the holomorphic Yukawa couplings is described by an Eisenstein series with weight 4: 
\begin{align}
    y_{111} &= 189\sum_{k=0}^\infty \frac{k^3 q^k_1}{1-q^k_1} - 27\sum_{m=0}^\infty \frac{(3m)^3 q^{3m}_1}{1-q^{3m}_1}
    \nonumber\\
    &= \frac{189}{240}\left( 1+ 240 \sum_{k=0}^\infty \frac{k^3 q^k_1}{1-q^k_1}\right) - \frac{27^2}{240} \left( 1 + 240 \sum_{m=0}^\infty \frac{(m)^3 q^{3m}}{1-q^{3m}_1}\right) + \frac{9}{4}\nonumber\\
    &= \frac{9}{4} + \frac{63}{80}E_4(t_1) - \frac{243}{80}E_4(3t_1),
    \nonumber\\
    y_{112} &= 3,\nonumber\\
    y_{122} &= 3,\nonumber\\
    y_{222} &= 0.
\end{align}
Since $E_4(3\tau)$ is not a modular form under $SL(2,\mathbb{Z})$, the effective action does not have the full  $SL(2,\mathbb{Z})$ modular symmetry. 
However, as shown in Appendix \ref{sec:app_weight4}, both $E_4(\tau)$ and $E_4(3\tau)$ are modular forms of weight 4 under $\Gamma_0(3)$. 
To see the modular invariance of the effective action, we move to the following basis:
\begin{align}
    t :=t_1,\qquad
    s := \frac{1}{2}t_1 + t_2,
\end{align}
where $q_2\ll q_1$ corresponds to $q_s \ll q_t$ with $q_s:=q_1^{1/2}q_2$ and $q_t:=q_1$. 
In this basis, the prepotential is rewritten in terms of the redefined moduli as
\begin{align}
    {\cal F} &= -\frac{1}{6}\left(-\frac{9}{4} t^3 + 9 t s^2 \right).
\end{align}
Since the moduli K\"ahler potential is approximately given in the regime ${\rm Im}(t)\ll {\rm Im}(s)$:
\begin{align}
K_{\rm ks} &= - \ln \biggl[\frac{3i}{2}(t - \widebar{t})(s- \widebar{s})^2\biggl],
\end{align}
the corresponding moduli K\"ahler metric is found as
\begin{align}
    &(K_{\rm ks})_{t\Bar{t}} \simeq - \frac{1}{(t-\widebar{t})^2},
    \qquad
    (K_{\rm ks})_{s\Bar{s}} \simeq - \frac{2}{(s-\widebar{s})^2},
    \qquad
    (K_{\rm ks})_{t\Bar{s}} = (K_{\rm ks})_{s\Bar{t}} \simeq 0. 
\end{align}
Thus, one can read off the matter modular weights from a $t$-dependent part of the moduli K\"ahler metric by using Eq. (\ref{eq:KahlerMetric}),  
\begin{align}
    K^{({\bf 27})}_{t\widebar{t}} &\sim e^{\frac{1}{3}(-K_{\rm ks})}(K_{\rm ks})_{t\widebar{t}} \propto (t-\widebar{t})^{-5/3},
    \nonumber\\
    K^{({\bf 27})}_{s\widebar{s}} &\sim e^{\frac{1}{3}(-K_{\rm ks})}(K_{\rm ks})_{s\widebar{s}} \propto (t-\widebar{t})^{1/3}.
\end{align}
It leads to the modular weight $-\frac{5}{3}$ for $A^{({\bf 27})}_t$ and $\frac{1}{3}$ for $A^{({\bf 27})}_s$. 
In this basis, the nonvanishing holomorphic Yukawa couplings are also rewritten as
\begin{align}
    y_{ttt} &= \frac{63}{80}E_4(t) - \frac{243}{80}E_4(3t)\qquad({\rm weight}\,4),\nonumber\\
    y_{tss} &= 9\hspace{118pt}({\rm weight}\,0),
\end{align}
which correspond to the modular forms of weight 4 and 0 under $\Gamma_0(3)$, respectively. 
Note that $y_{tts}$ vanishes, although the modular weight of $y_{tts}$ must be 2 and 
there exists a modular form of weight 2 under $\Gamma_0(3)$. 
(See Appendix \ref{sec:app}.)
It suggests that the low-energy effective field theory derived from this CY compactification 
has another coupling selection rule in addition to the $\Gamma_0(3)$ modular symmetry.
If $A^{({\bf 27})}_t$ has even $Z_2$ charge and  $A^{({\bf 27})}_s$ has odd $Z_2$ charge, 
the $y_{ttt}A^{({\bf 27})}_tA^{({\bf 27})}_tA^{({\bf 27})}_t$ coupling and the $y_{tss}A^{({\bf 27})}_t A^{({\bf 27})}_s A^{({\bf 27})}_s$ coupling are allowed, but the $y_{tts}A^{({\bf 27})}_t A^{({\bf 27})}_t A^{({\bf 27})}_s$ 
coupling is forbidden.
Such $Z_2$ charges can be originated from $Sp(6,\mathbb{Z})$.
At any rate, the above result is consistent with the modular invariance of 4D effective action.

To check the modular invariance of the effective action, let us consider the $\Gamma_0(3)$ modular transformation:
\begin{align}
    t \rightarrow \frac{at +b}{c t +d}
\end{align}
with $ad-bc=1$ and $c\equiv 0$ (mod $3$), the K\"ahler potential and superpotential 
\begin{align}
    K &= -\ln \biggl[\frac{i}{2}(t - \widebar{t})(s- \widebar{s})^2\biggl] +  K^{({\bf 27})}_{t\widebar{t}} |A^{({\bf 27})}_t|^2 
    + K^{({\bf 27})}_{s\widebar{s}} |A^{({\bf 27})}_s|^2,
    \nonumber\\
    W &= y_{ttt}A^{({\bf 27})}_tA^{({\bf 27})}_tA^{({\bf 27})}_t + y_{tss}A^{({\bf 27})}_t A^{({\bf 27})}_s A^{({\bf 27})}_s
\end{align}
transform as
\begin{align}
K \rightarrow K + \ln |ct +d|^2,\qquad
W \rightarrow (ct +d)^{-1}W,
\end{align}
and the K\"ahler invariant quantity $e^K |W|^2$ is modular invariant.

Note that instanton numbers $n_{0,d_2}$ have the same behavior as $n_{d_1,0}$.
Similarly,  we can discuss the regime $q_1 \ll q_2$.
Then, we can find the $\Gamma_0(3)$ modular symmetry and its modular forms.

\subsubsection{Three moduli}
\label{sec:gamma03_three}

We move to a favorable complete intersection CY threefold:
\begin{align}
\begin{matrix}
\mathbb{CP}^{1}\\
\mathbb{CP}^{1}\\
\mathbb{CP}^{3}\\
\end{matrix}
\begin{bmatrix}
0 & 2\\
2 & 0\\
3 & 1\\
\end{bmatrix}
,
\end{align} 
where the number of K\"ahler moduli is $h^{1,1}=3$. 
As studied in Ref.  \cite{Klemm:2004km}, the leading part of the prepotential for three K\"ahler moduli is calculated as
\begin{align}
    {\cal F} &= -\frac{1}{6}\left( 18t_1^2t_2  + 18 t_1^2t_3  + 18t_1t_2t_3\right),
\end{align}
where the triple intersection numbers are
\begin{align}
    \kappa_{112}=6, \quad
    \kappa_{113}=6,\quad
    \kappa_{123}=3,
\end{align}
and otherwise 0. The holomorphic Yukawa couplings of matter fields are determined by the classical part and instanton 
contribution as in Eq. (\ref{eq:Yukawa_3mod}). 
Let us consider the regime $q_2 \ll q_1, q_3$ such that only $q_{1,3}$ are relevant for the instanton expansion. 
The nonvanishing instanton numbers $n_{d_1, 0, d_3}$ are given in Table \ref{tab:instantons_gamma03_2}.
\begin{table}[H]
    \centering
    \begin{tabular}{c|c|c|c|c|c|}
       $d_1$ $\setminus$ $d_3$  &  0  & 1 & 2 & 3 & 4\\ \hline
       0  & & 18 &  &  & \\
       1  & 216 & 216 &  &  &  \\       
       2  & 216 & 2106 & 216 &  & \\       
       3  & 48 & 17856 & 17856 & 48 &  \\  
       4  & 216 & 95094 & 414720 & 95094  & 216\\ \hline     
    \end{tabular}
    \caption{Nonvanishing instanton numbers up to $d_1, d_2 \leq 4$.}
    \label{tab:instantons_gamma03_2}
\end{table}
As seen in the previous section, instanton numbers $n_{d_1,0,0}$ and $n_{d_1, 0, d_1}$ have an interesting structure such as 
$n_{d_1,0,0} = n_{d_1, 0, d_1}=48$ for $d_1 = 3 \mathbb{Z}_+$ and otherwise 216. 
Thus, when we restrict ourselves to the regime $q_3 \ll q_1$ such that the $t_3$-dependent instanton corrections are suppressed, only 
the modulus $t_1$ contributes to the holomorphic Yukawa couplings.\footnote{We can realize the same structure in the regime $q_1=q_3$.} 
In particular, one of the holomorphic Yukawa couplings is described by an Eisenstein series with weight 4. 
We list the nonvanishing holomorphic Yukawa couplings:
\begin{align}
    y_{111} &= 216\sum_{k=0}^\infty \frac{k^3 q_1^k}{1-q_1^k} - 168 \sum_{m=0}^\infty \frac{(3m)^3 q_1^{3m}}{1-q_1^{3m}}
    \nonumber\\
    &= \frac{216}{240}\left( 1+ 240 \sum_{k=0}^\infty \frac{k^3 q_1^k}{1-q_1^k}\right) - \frac{168\times 27}{240} \left( 1 + 240 \sum_{m=0}^\infty \frac{(m)^3 q_1^{3m}}{1-q_1^{3m}}\right) +18\nonumber\\
    &= 18 + \frac{9}{10}E_4(t_1) - \frac{189}{10}E_4(3t_1),
    \nonumber\\
    y_{112} &= 6,\nonumber\\
    y_{113} &= 6,\nonumber\\
    y_{123} &= 3.
\end{align}
Thus, we observe the $\Gamma_0(3)$ modular form in the same manner as in the previous example. 
To see the modular invariance of the effective action, we move to the following basis:
\begin{align}
    t :=t_1,\qquad
    s := t_1 + t_2,\qquad
    u : = t_1 + t_3,
\end{align}
where the regime $q_2 , q_3 \ll q_1$ correspond to $q_s, q_u \ll q_t$ with $q_u := e^{2\pi i u}$, $q_s := e^{2\pi i s}$ and $q_t := e^{2\pi i t}$. 
In this basis, the prepotential is rewritten in terms of the redefined moduli as
\begin{align}
    {\cal F} &= -\frac{1}{6}\left(-18 t^3 + 18 s t u \right).
\end{align}
Since the moduli K\"ahler potential is approximately given in the regime ${\rm Im}(t)\ll {\rm Im}(s),{\rm Im}(u)$:
\begin{align}
K_{\rm ks} &= - \ln \biggl[3i(t - \widebar{t})(s- \widebar{s})(u- \widebar{u})\biggl],
\end{align}
the corresponding moduli K\"ahler metric is found as
\begin{align}
    &(K_{\rm ks})_{t\Bar{t}} \simeq - \frac{1}{(t-\widebar{t})^2},
    \qquad
    (K_{\rm ks})_{s\Bar{s}} \simeq - \frac{1}{(s-\widebar{s})^2},
    \qquad
    (K_{\rm ks})_{u\Bar{u}} \simeq - \frac{1}{(u-\widebar{u})^2}, 
\end{align}
and otherwise 0. 
Thus, one can read off the matter modular weights from a $t$-dependent part of the moduli K\"ahler metric by using Eq. (\ref{eq:KahlerMetric}),  
\begin{align}
    K^{({\bf 27})}_{t\widebar{t}} &\sim e^{\frac{1}{3}(-K_{\rm ks})}(K_{\rm ks})_{t\widebar{t}} \propto (t-\widebar{t})^{-5/3},
    \nonumber\\
    K^{({\bf 27})}_{s\widebar{s}} &\sim e^{\frac{1}{3}(-K_{\rm ks})}(K_{\rm ks})_{s\widebar{s}} \propto (t-\widebar{t})^{1/3},
    \nonumber\\
    K^{({\bf 27})}_{u\widebar{u}} &\sim e^{\frac{1}{3}(-K_{\rm ks})}(K_{\rm ks})_{u\widebar{u}} \propto (t-\widebar{t})^{1/3}.
\end{align}
It leads to the modular weight $-\frac{5}{3}$ for $A^{({\bf 27})}_t$ and $\frac{1}{3}$ for $A^{({\bf 27})}_s$ and $A^{({\bf 27})}_u$. 
The holomorphic Yukawa couplings are also rewritten as
\begin{align}
    y_{ttt} &= \frac{9}{10}E_4(t) - \frac{189}{10}E_4(3t)\qquad({\rm weight}\,4),\nonumber\\
    y_{tsu} &= 3\hspace{118pt}({\rm weight}\,0),
\end{align}
and otherwise 0. 
In this basis, they are described by modular forms of weights 4 and 0 under $\Gamma_0(3)$, respectively. 
Thus, this result is consistent with the modular invariance of 4D effective action. Note that the effective theory has the same coupling selection rule, as shown in Sec. \ref{sec:sl2_three}.

To check the modular invariance of the effective action, let us consider the $\Gamma_0(3)$ modular transformation:
\begin{align}
    t \rightarrow \frac{at +b}{c t +d}
\end{align}
with $ad-bc=1$ and $c\equiv 0$ (mod $3$), under which the K\"ahler potential and superpotential 
\begin{align}
    K &= - \ln \biggl[i (t - \widebar{t})(u- \widebar{u})(s- \widebar{s})\biggl] +  \sum_{a,b} K^{({\bf 27})}_{a\widebar{b}} A^{({\bf 27})}_a \widebar{A^{({\bf 27})}_b},
    \nonumber\\
    W &= y_{ttt}A^{({\bf 27})}_tA^{({\bf 27})}_tA^{({\bf 27})}_t 
    + y_{stu}A^{({\bf 27})}_s A^{({\bf 27})}_t A^{({\bf 27})}_u
\end{align}
transform as
\begin{align}
K \rightarrow K + \ln |ct +d|^2,\qquad
W \rightarrow (ct +d)^{-1}W.
\end{align}
Thus, the K\"ahler invariant quantity $e^K |W|^2$ is modular invariant.

\subsection{$\Gamma_0(4)$ modular forms}
\label{sec:gamma04}

In this section, we focus on specific CY threefolds with three moduli in Sec. \ref{sec:gamma04_three} and four moduli 
in Sec. \ref{sec:gamma04_four}, leading to $\Gamma_0(4)$ modular forms for holomorphic Yukawa couplings. 
The $\Gamma_0(N)$ modular forms are summarized in Appendix \ref{sec:app}.

\subsubsection{Three moduli}
\label{sec:gamma04_three}

We start with a CY threefold defined by a single polynomial in three projective spaces:
\begin{align}
\begin{matrix}
\mathbb{CP}^{1}\\
\mathbb{CP}^{1}\\
\mathbb{CP}^{3}\\
\end{matrix}
\begin{bmatrix}
2\\
2\\
3\\
\end{bmatrix}
,
\end{align} 
where the number of K\"ahler moduli is $h^{1,1}=3$. 
As studied in Ref.~\cite{Berglund:1993ax}, the leading part of the prepotential for three K\"ahler moduli is calculated as
\begin{align}
    {\cal F} &= -\frac{1}{6}\left( 6t_1t_3^2  + 6t_2t_3^2  + 18t_1t_2t_3\right),
\end{align}
where the triple intersection numbers are
\begin{align}
    \kappa_{233}=2, \quad
    \kappa_{133}=2,\quad
    \kappa_{123}=3,
\end{align}
and otherwise 0. The holomorphic Yukawa couplings of matter fields are determined by the classical part and instanton 
contribution as in Eq. (\ref{eq:Yukawa_3mod}). 
Let us consider the regime $q_3 \ll q_1, q_2$; only $q_{1,2}$ contribute to the instanton expansion. 
The nonvanishing instanton numbers $n_{d_1, d_2, 0}$ are given in Table \ref{tab:instantons_gamma04_1}.
\begin{table}[H]
    \centering
    \begin{tabular}{c|c|c|c|c|c|}
       $d_1$ $\setminus$ $d_2$  &  0  & 1 & 2 & 3 & 4\\ \hline
       0  & & 54 &   &  & \\
       1  & 54 & 180 & 54 &  &  \\       
       2  &  & 54 & 144 & 54 & \\       
       3  &  &  & 54 & 180 & 54 \\  
       4  &  &  &  & 54  & 144\\ \hline     
    \end{tabular}
    \caption{Nonvanishing instanton numbers up to $d_1, d_2 \leq 4$.}
    \label{tab:instantons_gamma04_1}
\end{table}
Remarkably, instanton numbers $n_{d_1,0,0}$ and $n_{d_1, d_2, 0}$ have an interesting structure, i.e., 
$n_{d_1+1,d_1,0}=n_{d_1, d_1+1,0}=54$ for all $d_1$, and $n_{d_1,d_1} = 144$ for $d_1 = 2 \mathbb{Z}_+$ and otherwise 180. 
Specifically, nonvanishing holomorphic Yukawa couplings are described by 
\begin{align}
y_{111} &= \sum_{d_1,d_2=0}^\infty \frac{(d_1)^3n_{d_1, d_2, 0}q_1^{d_1}q_2^{d_2}}{1-q_1^{d_1}q_2^{d_2}}
\nonumber\\
&= \sum_{d_1=0}^\infty \biggl[\frac{(d_1)^3n_{d_1, d_1, 0}q_1^{d_1}q_2^{d_1}}{1-q_1^{d_1}q_2^{d_1}}
+\frac{(d_1)^3n_{d_1, d_1+1, 0}q_1^{d_1}q_2^{d_1+1}}{1-q_1^{d_1}q_2^{d_1+1}}
+\frac{(d_1+1)^3n_{d_1+1, d_1, 0}q_1^{d_1+1}q_2^{d_1}}{1-q_1^{d_1+1}q_2^{d_1}}\biggl],
\nonumber\\
y_{112} &= \sum_{d_1,d_2=0}^\infty \frac{(d_1)^2d_2 n_{d_1, d_2, 0}q_1^{d_1}q_2^{d_2}}{1-q_1^{d_1}q_2^{d_2}}
\nonumber\\
&= \sum_{d_1=0}^\infty \biggl[\frac{(d_1)^3n_{d_1, d_1, 0}q_1^{d_1}q_2^{d_1}}{1-q_1^{d_1}q_2^{d_1}}
+\frac{(d_1)^2(d_1+1)n_{d_1, d_1+1, 0}q_1^{d_1}q_2^{d_1+1}}{1-q_1^{d_1}q_2^{d_1+1}}
+\frac{(d_1+1)^2d_1 n_{d_1+1, d_1, 0}q_1^{d_1+1}q_2^{d_1}}{1-q_1^{d_1+1}q_2^{d_1}}\biggl],
\nonumber\\
y_{122} &= \sum_{d_1,d_2=0}^\infty \frac{d_1(d_2)^2 n_{d_1, d_2, 0}q_1^{d_1}q_2^{d_2}}{1-q_1^{d_1}q_2^{d_2}}
\nonumber\\
&= \sum_{d_1=0}^\infty \biggl[\frac{(d_1)^3n_{d_1, d_1, 0}q_1^{d_1}q_2^{d_1}}{1-q_1^{d_1}q_2^{d_1}}
+\frac{d_1(d_1+1)^2n_{d_1, d_1+1, 0}q_1^{d_1}q_2^{d_1+1}}{1-q_1^{d_1}q_2^{d_1+1}}
+\frac{(d_1+1)(d_1)^2 n_{d_1+1, d_1, 0}q_1^{d_1+1}q_2^{d_1}}{1-q_1^{d_1+1}q_2^{d_1}}\biggl]
\nonumber\\
&= y_{112},
\nonumber\\
y_{222} &= \sum_{d_1,d_2=0}^\infty \frac{(d_2)^3 n_{d_1, d_2, 0}q_1^{d_1}q_2^{d_2}}{1-q_1^{d_1}q_2^{d_2}}
\nonumber\\
&= \sum_{d_1=0}^\infty \biggl[\frac{(d_1)^3n_{d_1, d_1, 0}q_1^{d_1}q_2^{d_1}}{1-q_1^{d_1}q_2^{d_1}}
+\frac{(d_1+1)^3 n_{d_1, d_1+1, 0}q_1^{d_1}q_2^{d_1+1}}{1-q_1^{d_1}q_2^{d_1+1}}
+\frac{(d_1)^3 n_{d_1+1, d_1, 0}q_1^{d_1+1}q_2^{d_1}}{1-q_1^{d_1+1}q_2^{d_1}}\biggl]
\nonumber\\
&= y_{111},
\nonumber\\
y_{123}&=3,\qquad
y_{233}=y_{133}=2.
\end{align}
When we restrict ourselves to the regime $t_1 =t_2$, 
they are rewritten as
\begin{align}
y_{111} 
&= \sum_{d_1=0}^\infty \biggl[\frac{(d_1)^3n_{d_1, d_1, 0}q_1^{2d_1}}{1-q_1^{2d_1}}
+54\frac{(d_1)^3q_1^{2d_1+1}}{1-q_1^{2d_1+1}}
+54\frac{(d_1+1)^3 q_1^{2d_1+1}}{1-q_1^{2d_1+1}}\biggl]
\nonumber\\
&=\sum_{d_1=0}^\infty \biggl[180\frac{(d_1)^3q_1^{2d_1}}{1-q_1^{2d_1}}
-36\frac{(2d_1)^3q_1^{2(2d_1)}}{1-q_1^{2(2d_1)}}
+\frac{54}{4}\left(\frac{(2d_1+1)^3q_1^{2d_1+1}}{1-q_1^{2d_1+1}}
+3\frac{(2d_1+1)q_1^{2d_1+1}}{1-q_1^{2d_1+1}}\right)\biggl]
\nonumber\\
&=180\left(\frac{E_4(2t_1)-1}{240}\right) - 288\left(\frac{E_4(4t_1)-1}{240}\right)
\nonumber\\
&+\frac{54}{4}\sum_{d_1=0}^\infty \Biggl[ 
\left(\frac{(d_1)^3 q_1^{d_1}}{1-q_1^{d_1}} - \frac{(2d_1)^3 q_1^{2d_1}}{1-q_1^{2d_1}}\right)
+3\left(\frac{(d_1) q_1^{d_1}}{1-q_1^{d_1}} - \frac{(2d_1)q_1^{2d_1}}{1-q_1^{2d_1}}\right)\Biggl]
\nonumber\\
&=-\frac{27}{32} + \frac{3}{160}(3 E_4(t_1) + 16\left( E_4(2t_1) - 4 E_4(4t_1)\right) 
- \frac{27}{16}\left(E_2(t_1)  - 2E_2(2t_1)\right),
\nonumber\\
y_{112} 
&= \sum_{d_1=0}^\infty \biggl[\frac{(d_1)^3n_{d_1, d_1, 0}q_1^{2d_1}}{1-q_1^{2d_1}}
+54\frac{(d_1)^2(d_1+1)q_1^{2d_1+1}}{1-q_1^{2d_1+1}}
+54\frac{d_1(d_1+1)^2 q_1^{2d_1+1}}{1-q_1^{2d_1+1}}\biggl]
\nonumber\\
&=180\left(\frac{E_4(2t_1)-1}{240}\right) - 288\left(\frac{E_4(4t_1)-1}{240}\right)
+\frac{27}{2}\sum_{d_1=0}^\infty \Biggl[ 
\left(\frac{(2d_1+1)^3q_1^{2d_1+1}}{1-q_1^{2d_1+1}}
-\frac{(2d_1+1)q_1^{2d_1+1}}{1-q_1^{2d_1+1}}\right)\Biggl]
\nonumber\\
&=\frac{45}{32} + \frac{3}{160}(3 E_4(t_1) + 16\left( E_4(2t_1) - 4 E_4(4t_1)\right) 
+ \frac{9}{16}\left(E_2(t_1)  - 2E_2(2t_1)\right),
\nonumber\\
y_{122} &= y_{112},
\nonumber\\
y_{222} &= y_{111},
\nonumber\\
y_{123}&=3,\qquad
y_{233}=y_{133}=2.
\end{align}
Note that $E_2(t_1)  - 2E_2(2t_1)$ and $E_4(nt_1)$ for $n|N$ are modular forms of weight 2 and 4 under $\Gamma_0(4)$, 
respectively. See Appendix \ref{sec:app}, for more details about the modular form of $\Gamma_0(N)$. 
To see the modular invariance of the effective action, we move to the following basis:
\begin{align}
    t := \frac{t^1-t^2}{2},\qquad
    s := \frac{t^1 + t^2}{2},\qquad
    u : = 2t_3 + \frac{3}{4}(t^1 + t^2).
\end{align}
In this basis, the prepotential is rewritten in terms of the redefined moduli as
\begin{align}
    {\cal F} &= -\frac{1}{6}\left(-\frac{27}{4}s^3 +\frac{27}{2} st^2 - 9 t^2u +3su^2\right).
\end{align}
The corresponding moduli K\"ahler potential is given by
\begin{align}
K_{\rm ks} &= - \ln \biggl[\frac{i}{2}(s- \widebar{s})(u- \widebar{u})^2-\frac{9i}{8}(s - \widebar{s})^3 + \frac{9i}{4}(s- \widebar{s})(t- \widebar{t})^2 - \frac{3i}{2}(t- \widebar{t})^2(u- \widebar{u})\biggl].
\end{align}
In an asymptotic regime $q_u \ll q_s\ll q_t$ with $q_t := e^{2\pi i t}$, $q_u := e^{2\pi i u}$ and $q_s := e^{2\pi i s}$, 
corresponding to $q_3\ll q_1,q_{2}$ and $q_2 \rightarrow q_1$, 
the moduli K\"ahler metric is obtained as
\begin{align}
    &(K_{\rm ks})_{s\Bar{s}} \simeq - \frac{1}{(s-\widebar{s})^2},
    \qquad
    (K_{\rm ks})_{u\Bar{u}} \simeq - \frac{2}{(u-\widebar{u})^2},
    \qquad
    (K_{\rm ks})_{t\Bar{t}} \simeq - \frac{3}{2(s-\widebar{s})(u-\widebar{u})}, 
\end{align}
and otherwise 0. 
Thus, one can read off the matter modular weights from a $s$-dependent part of the moduli K\"ahler metric by using Eq. (\ref{eq:KahlerMetric}),  
\begin{align}
    K^{({\bf 27})}_{s\widebar{s}} &\sim e^{\frac{1}{3}(-K_{\rm ks})}(K_{\rm ks})_{s\widebar{s}} \propto (s-\widebar{s})^{-5/3},
    \nonumber\\
    K^{({\bf 27})}_{u\widebar{u}} &\sim e^{\frac{1}{3}(-K_{\rm ks})}(K_{\rm ks})_{u\widebar{u}} \propto (s-\widebar{s})^{1/3},
    \nonumber\\
    K^{({\bf 27})}_{t\widebar{t}} &\sim e^{\frac{1}{3}(-K_{\rm ks})}(K_{\rm ks})_{t\widebar{t}} \propto (s-\widebar{s})^{-2/3}.
\end{align}
It leads to the modular weight $-\frac{5}{3}$ for $A^{({\bf 27})}_s$, $\frac{1}{3}$ for $A^{({\bf 27})}_u$ and $-\frac{2}{3}$ for $A^{({\bf 27})}_t$. Then, the modular weights of holomorphic Yukawa couplings are determined, and their explicit forms are rewritten as
\begin{align}
    y_{sss} &= \frac{9}{20} E_4(s) + \frac{12}{5}\left( E_4(2s) - 4 E_4(4s)\right)\quad ({\rm weight}\,4),\nonumber\\
    y_{stt} &= -\frac{9}{2}\left(E_2(s)  - 2E_2(2s)\right)\hspace{65pt} ({\rm weight}\,2),\nonumber\\
    y_{suu} &= 1\hspace{165pt} ({\rm weight}\,0),
\end{align}
and otherwise 0. 
Note that modular forms of the weight 2 under $\Gamma_0(4)$ are given by a linear combination: $E_2(s)-2E_2(2s)$ and $E_2(2s)- 2 E_2(4s)$, as shown in Appendix \ref{sec:app_weight2}.

To check the modular invariance of the effective action, let us consider the $\Gamma_0(4)$ modular transformation:
\begin{align}
    s \rightarrow \frac{as +b}{c s +d}
\end{align}
with $ad-bc=1$ and $c\equiv 0$ (mod $4$), under which the K\"ahler potential and superpotential 
\begin{align}
    K &= - \ln \biggl[\frac{i}{2}(s- \widebar{s})(u- \widebar{u})^2\biggl] +  \sum_{a,b} K^{({\bf 27})}_{a\widebar{b}} A^{({\bf 27})}_a \widebar{A^{({\bf 27})}_b},
    \nonumber\\
    W &= y_{sss}A^{({\bf 27})}_sA^{({\bf 27})}_sA^{({\bf 27})}_s + y_{stt}A^{({\bf 27})}_s A^{({\bf 27})}_t A^{({\bf 27})}_t + y_{suu}A^{({\bf 27})}_s A^{({\bf 27})}_u A^{({\bf 27})}_u 
    + y_{ttu}A^{({\bf 27})}_t A^{({\bf 27})}_t A^{({\bf 27})}_u 
\end{align}
transform as
\begin{align}
K \rightarrow K + \ln |cs +d|^2,\qquad
W \rightarrow (cs +d)^{-1}W.
\end{align}
Thus, the K\"ahler invariant quantity $e^K |W|^2$ is modular invariant.

\subsubsection{Four moduli}
\label{sec:gamma04_four}

We next discuss a CY threefold defined by a single polynomial in four projective spaces:
\begin{align}
\begin{matrix}
\mathbb{CP}^{1}\\
\mathbb{CP}^{1}\\
\mathbb{CP}^{1}\\
\mathbb{CP}^{1}\\
\end{matrix}
\begin{bmatrix}
2\\
2\\
2\\
2\\
\end{bmatrix}
,
\end{align} 
where the number of K\"ahler moduli is $h^{1,1}=4$. 
As studied in heterotic string theory with line bundles \cite{Buchbinder:2013dna}, the leading part of the prepotential for four K\"ahler moduli is calculated as
\begin{align}
    {\cal F} &= -\frac{1}{6}\left( 12t_1t_2t_3  + 12t_1t_2t_4  + 12t_1t_3t_4 + 12t_2t_3t_4\right),
\end{align}
where the triple intersection numbers are
\begin{align}
    \kappa_{123}= \kappa_{124}=\kappa_{134}=\kappa_{234}=2,
\end{align}
and otherwise 0. The holomorphic Yukawa couplings of matter fields are determined by the classical part and instanton 
contribution:
\begin{align}
    y_{abc} = \kappa_{abc} + \sum_{d_1, d_2, d_3,d_4=0}^\infty \frac{c_{abc}(d_1,d_2,d_3,d_4)n_{d_1, d_2, d_3,d_4}q_1^{d_1} q_2^{d_2}q_3^{d_3}q_4^{d_4}}{1-q_1^{d_1} q_2^{d_2}q_3^{d_3}q_4^{d_4}}
    \label{eq:Yukawa_4mod}
\end{align}
with
\begin{align}
    c_{abc}(d_1,d_2,d_3) = d_a d_b d_c,\quad
    q_1 = e^{2\pi i t_1},\quad 
    q_2 = e^{2\pi i t_2},\quad
    q_3 = e^{2\pi i t_3},\quad
    q_4 = e^{2\pi i t_4}.
\end{align}
To simplify our analysis, let us focus on the regime $q_3, q_4 \ll q_1, q_2$; only $q_{1,2}$ contribute to the instanton expansion. 
The nonvanishing instanton numbers $n_{d_1, d_2, 0,0}$ are given in Table \ref{tab:instantons_gamma04_2}.
\begin{table}[H]
    \centering
    \begin{tabular}{c|c|c|c|c|c|}
       $d_1$ $\setminus$ $d_2$  &  0  & 1 & 2 & 3 & 4\\ \hline
       0  & & 48 &   &  & \\
       1  & 48 & 160 & 48 &  &  \\       
       2  &  & 48 & 128 & 48 & \\       
       3  &  &  & 48 & 160 & 48 \\  
       4  &  &  &  & 48  & 128\\ \hline     
    \end{tabular}
    \caption{Nonvanishing instanton numbers up to $d_1, d_2 \leq 4$.}
    \label{tab:instantons_gamma04_2}
\end{table}
As shown in the previous example, instanton numbers have an interesting structure, i.e., 
$n_{d_1+1,d_1,0,0}=n_{d_1, d_1+1,0,0}=48$ for all $d_1$, and $n_{d_1,d_1,0,0} = 128$ for $d_1 = 2 \mathbb{Z}_+$ and otherwise 160. 
In a similar manner as in Sec. \ref{sec:gamma04_three}, we restrict ourselves to the regime $t_1 =t_2$ in which nonvanishing holomorphic Yukawa couplings are described by 
\begin{align}
y_{111} 
&= \sum_{d_1=0}^\infty \biggl[\frac{(d_1)^3n_{d_1, d_1, 0, 0}q_1^{2d_1}}{1-q_1^{2d_1}}
+48\frac{(d_1)^3q_1^{2d_1+1}}{1-q_1^{2d_1+1}}
+48\frac{(d_1+1)^3 q_1^{2d_1+1}}{1-q_1^{2d_1+1}}\biggl]
\nonumber\\
&=\sum_{d_1=0}^\infty \biggl[160\frac{(d_1)^3q_1^{2d_1}}{1-q_1^{2d_1}}
-32\frac{(2d_1)^3q_1^{2(2d_1)}}{1-q_1^{2(2d_1)}}
+\frac{48}{4}\left(\frac{(2d_1+1)^3q_1^{2d_1+1}}{1-q_1^{2d_1+1}}
+3\frac{(2d_1+1)q_1^{2d_1+1}}{1-q_1^{2d_1+1}}\right)\biggl]
\nonumber\\
&=160\left(\frac{E_4(2t_1)-1}{240}\right) - 256\left(\frac{E_4(4t_1)-1}{240}\right)
\nonumber\\
&+\frac{48}{4}\sum_{d_1=0}^\infty \Biggl[ 
\left(\frac{(d_1)^3 q_1^{d_1}}{1-q_1^{d_1}} - \frac{(2d_1)^3 q_1^{2d_1}}{1-q_1^{2d_1}}\right)
+3\left(\frac{(d_1) q_1^{d_1}}{1-q_1^{d_1}} - \frac{(2d_1)q_1^{2d_1}}{1-q_1^{2d_1}}\right)\Biggl]
\nonumber\\
&=-\frac{3}{4} + \frac{1}{60}(3 E_4(t_1) + 16\left( E_4(2t_1) - 4 E_4(4t_1)\right) 
- \frac{3}{2}\left(E_2(t_1)  - 2E_2(2t_1)\right),
\nonumber\\
y_{112} 
&= \sum_{d_1=0}^\infty \biggl[\frac{(d_1)^3n_{d_1, d_1, 0}q_1^{2d_1}}{1-q_1^{2d_1}}
+54\frac{(d_1)^2(d_1+1)q_1^{2d_1+1}}{1-q_1^{2d_1+1}}
+54\frac{d_1(d_1+1)^2 q_1^{2d_1+1}}{1-q_1^{2d_1+1}}\biggl]
\nonumber\\
&=160\left(\frac{E_4(2t_1)-1}{240}\right) - 256\left(\frac{E_4(4t_1)-1}{240}\right)
+12\sum_{d_1=0}^\infty \Biggl[ 
\left(\frac{(2d_1+1)^3q_1^{2d_1+1}}{1-q_1^{2d_1+1}}
-\frac{(2d_1+1)q_1^{2d_1+1}}{1-q_1^{2d_1+1}}\right)\Biggl]
\nonumber\\
&=\frac{5}{4} + \frac{1}{60}(3 E_4(t_1) + 16\left( E_4(2t_1) - 4 E_4(4t_1)\right) 
+ \frac{1}{2}\left(E_2(t_1)  - 2E_2(2t_1)\right),
\nonumber\\
y_{122} &= y_{112},
\nonumber\\
y_{222} &= y_{111},
\nonumber\\
y_{123}&=y_{124}=y_{133}=y_{234}=2.
\end{align}
Note that $E_2(t_1)  - 2E_2(2t_1)$ and $E_4(nt_1)$ for $n|N$ are modular forms of weight 2 and 4 under $\Gamma_0(4)$, 
respectively. 
To see the modular invariance of the effective action, we move to the following basis:
\begin{align}
    t := \frac{t^1-t^2}{2},\qquad
    s := \frac{t^1 + t^2}{2},\qquad
    u : = t_3 + \frac{1}{4}(t^1 + t^2),\qquad
    r : = t_4 + \frac{1}{4}(t^1 + t^2).
\end{align}
In this basis, the prepotential is rewritten in terms of the redefined moduli as
\begin{align}
    {\cal F} &= -\frac{1}{6}\left(24sur -6s^3  -12t^2u -12t^2r + 12t^2s\right).
\end{align}
The corresponding moduli K\"ahler potential is approximately given by
\begin{align}
K_{\rm ks} &= - \ln \biggl[i\left\{4(s- \widebar{s})(u- \widebar{u})(r- \widebar{r})-(s - \widebar{s})^3 -2(t- \widebar{t})^2\left((r- \widebar{r}) +(t- \widebar{t})\right) -(s- \widebar{s})\right\} \biggl].
\end{align}
Note that we consider an asymptotic regime $q_3, q_4 \ll q_1, q_2$ with $q_1 = q_2$, 
corresponding to $q_u, q_r \ll q_s\ll q_t$ with $q_t := e^{2\pi i t}=1$, $q_s := e^{2\pi i s}$, $q_u := e^{2\pi i u}$ and $q_r := e^{2\pi i r}$, 
the corresponding moduli K\"ahler metric is found as
\begin{align}
    &(K_{\rm ks})_{s\Bar{s}} \simeq - \frac{1}{(s-\widebar{s})^2},
    \qquad
    (K_{\rm ks})_{t\Bar{t}} \simeq - \frac{1}{(s-\widebar{s})}\left(\frac{1}{u-\widebar{u}} + \frac{1}{r-\widebar{r}}\right), 
    \nonumber\\
 &(K_{\rm ks})_{u\Bar{u}} \simeq - \frac{1}{(u-\widebar{u})^2},\qquad 
     (K_{\rm ks})_{r\Bar{r}} \simeq - \frac{1}{(r-\widebar{r})^2}, 
\end{align}
and otherwise 0. 
Thus, one can read off the matter modular weights from a $s$-dependent part of the moduli K\"ahler metric by using Eq. (\ref{eq:KahlerMetric}),  
\begin{align}
    K^{({\bf 27})}_{s\widebar{s}} &\sim e^{\frac{1}{3}(-K_{\rm ks})}(K_{\rm ks})_{s\widebar{s}} \propto (s-\widebar{s})^{-5/3},
\nonumber\\K^{({\bf 27})}_{t\widebar{t}} &\sim e^{\frac{1}{3}(-K_{\rm ks})}(K_{\rm ks})_{t\widebar{t}} \propto (s-\widebar{s})^{-2/3},
    \nonumber\\
    K^{({\bf 27})}_{u\widebar{u}} &\sim e^{\frac{1}{3}(-K_{\rm ks})}(K_{\rm ks})_{u\widebar{u}} \propto (s-\widebar{s})^{1/3},
    \nonumber\\
    K^{({\bf 27})}_{r\widebar{r}} &\sim e^{\frac{1}{3}(-K_{\rm ks})}(K_{\rm ks})_{r\widebar{r}} \propto (s-\widebar{s})^{1/3}.
\end{align}
It leads to the modular weight $-\frac{5}{3}$ for $A^{({\bf 27})}_s$, $-\frac{2}{3}$ for $A^{({\bf 27})}_t$, and $\frac{1}{3}$ for $A^{({\bf 27})}_u$ and $A^{({\bf 27})}_r$. Then, the modular weights of holomorphic Yukawa couplings are determined, and their explicit forms are rewritten as
\begin{align}
    y_{sss} &= \frac{6}{15} E_4(s) + \frac{32}{15}\left( E_4(2s) - 4 E_4(4s)\right)\quad ({\rm weight}\,4),\nonumber\\
    y_{stt} &= -4\left(E_2(s)  - 2E_2(2s)\right)\hspace{69pt} ({\rm weight}\,2),\nonumber\\
    y_{ttu} &= y_{ttr}=-4\hspace{127pt} ({\rm weight}\,0),\nonumber\\
    y_{sur} &=4\hspace{165pt} ({\rm weight}\,0),
\end{align}
and otherwise 0. Note that modular forms of the weight 2 under $\Gamma_0(4)$ are given by a linear combination: $E_2(s)-2E_2(2s)$ and $E_2(2s)- 2 E_2(4s)$, as shown in Appendix \ref{sec:app_weight2}.

To check the modular invariance of the effective action, let us consider the $\Gamma_0(4)$ modular transformation:
\begin{align}
    s \rightarrow \frac{as +b}{c s +d}
\end{align}
with $ad-bc=1$ and $c\equiv 0$ (mod $4$), under which the K\"ahler potential and superpotential 
\begin{align}
    K &= - \ln \biggl[4i(s- \widebar{s})(u- \widebar{u})(r- \widebar{r})\biggl] +  \sum_{a,b} K^{({\bf 27})}_{a\widebar{b}} A^{({\bf 27})}_a \widebar{A^{({\bf 27})}_b},
    \nonumber\\
    W &= y_{sss}A^{({\bf 27})}_sA^{({\bf 27})}_sA^{({\bf 27})}_s + y_{stt}A^{({\bf 27})}_s A^{({\bf 27})}_t A^{({\bf 27})}_t 
    \nonumber\\
    &+ y_{sur}A^{({\bf 27})}_s A^{({\bf 27})}_u A^{({\bf 27})}_r 
    + y_{ttr}A^{({\bf 27})}_t A^{({\bf 27})}_t A^{({\bf 27})}_r 
    + y_{ttu}A^{({\bf 27})}_t A^{({\bf 27})}_t A^{({\bf 27})}_u 
\end{align}
transform as
\begin{align}
K \rightarrow K + \ln |cs +d|^2,\qquad
W \rightarrow (cs +d)^{-1}W.
\end{align}
Thus, the K\"ahler invariant quantity $e^K |W|^2$ is modular invariant.

Note that the 4D low-energy effective theory has the permutation symmetry between 
$A^{({\bf 27})}_r$ and $A^{({\bf 27})}_u$.
Such a $Z_2$ permutation symmetry is originated from $Sp(10,\mathbb{Z})$.

\section{Hierarchical structure of physical Yukawa couplings}
\label{sec:hierarchy}

In this section, we discuss the flavor structure of physical Yukawa couplings in more details. 
Specifically, we focus on the CY threefold in Sec. \ref{sec:sl2_three}. 

To see the flavor structure of ${\bf 27}$ matters, let us consider the physical Yukawa couplings 
by diagonalizing the kinetic terms of ${\bf 27}$ matter fields:
\begin{align}
A^a K_{a\bar{b}}^{({\bf 27})} \bar{A}^{\bar{b}} = A^a (L^\dagger)^{\bar{\hat{c}}}_a \Lambda_{\bar{\hat{c}} \hat{c}} L^{\hat{c}}_{\bar{b}} \bar{A}^{\bar{b}} 
=: {\cal A}^{\hat{a}}\bar{{\cal A}}^{\bar{\hat{a}}}  
\end{align}
with 
\begin{align}
    {\cal A}^{\hat{a}} = A^d (L^\dagger)^{\bar{\hat{a}}}_{d} (\sqrt{\Lambda})_{\bar{\hat{a}}\hat{a}},
\end{align}
where $L^{\hat{c}}_{\bar{b}}$ is an unitary matrix to diagonalize the matter field K\"ahler metric $K_{a\bar{b}}^{({\bf 27})}$, and $\Lambda_{\bar{\hat{a}}\hat{a}}$ denotes its eigenvalue matrix, that is, 
$(L K L^\dagger)_{\bar{\hat{a}}\hat{a}} = \Lambda_{\bar{\hat{a}}\hat{a}}$. 
Then, the physical Yukawa couplings are expressed by
\begin{align}
    Y_{\hat{a}\hat{b}\hat{c}} = e^{K/2} (\Lambda^{-1/2}L)^d_{\hat{a}}(\Lambda^{-1/2}L)^e_{\hat{b}}(\Lambda^{-1/2}L)^f_{\hat{c}}y_{def},
\end{align}
where we use
\begin{align}
    e^{K/2} y_{def}A^dA^e A^f = e^{K/2}y_{def}(\Lambda^{-1/2}L)^d_{\hat{a}}(\Lambda^{-1/2}L)^e_{\hat{b}}(\Lambda^{-1/2}L)^f_{\hat{c}} {\cal A}^{\hat{a}}{\cal A}^{\hat{b}}{\cal A}^{\hat{c}} 
    = Y_{\hat{a}\hat{b}\hat{c}}{\cal A}^{\hat{a}}{\cal A}^{\hat{b}}{\cal A}^{\hat{c}}. 
\end{align}

We can estimate the physical Yukawa couplings in the regime ${\rm Im}(t_1) \ll {\rm Im}(t_2)={\rm Im}(t_3)$ by 
using the nonvanishing holomorphic Yukawa couplings (\ref{eq:hol_yukawa}) and the matter field K\"ahler metric:
\begin{align}
\begin{pmatrix}
    K_{1\Bar{1}} & K_{1\Bar{2}} & K_{1\Bar{3}}\\
    K_{2\Bar{1}} & K_{2\Bar{2}} & K_{2\Bar{3}}\\
    K_{3\Bar{1}} & K_{3\Bar{2}} & K_{3\Bar{3}}
\end{pmatrix}
= \frac{e^{\frac{K_{\rm cs}}{3}}}{2}
\begin{pmatrix}
    \frac{{\rm Im}(t_3)^{2/3}}{{\rm Im}(t_1)^{5/3}} & \frac{{\rm Im}(t_1)^{1/3}}{{\rm Im}(t_3)^{4/3}} & \frac{{\rm Im}(t_1)^{1/3}}{{\rm Im}(t_3)^{4/3}}\\
    \frac{{\rm Im}(t_1)^{1/3}}{{\rm Im}(t_3)^{4/3}} & \frac{{\rm Im}(t_1)^{1/3}}{{\rm Im}(t_3)^{4/3}} & 0\\    
    \frac{{\rm Im}(t_1)^{1/3}}{{\rm Im}(t_3)^{4/3}} & 0 & \frac{{\rm Im}(t_1)^{1/3}}{{\rm Im}(t_3)^{4/3}}\\    
\end{pmatrix}
.
\end{align}
Since the eigenvalues of the matter field K\"ahler metric 
\begin{align}
    \Lambda_{\bar{\hat{c}} \hat{c}} = \frac{e^{\frac{K_{\rm cs}}{3}}}{2} 
    {\rm diag}\left( \frac{{\rm Im}(t_1)^{1/3}}{{\rm Im}(t_3)^{4/3}},\,\, 
    \frac{{\rm Im}(t_1)^{1/3}}{{\rm Im}(t_3)^{4/3}},\,\, 
    \frac{{\rm Im}(t_3)^{2/3}}{{\rm Im}(t_1)^{5/3}}\right)
\end{align}
are hierarchical in the regime ${\rm Im}(t_1) \ll {\rm Im}(t_3)$, it will be accessible to the hierarchical physical Yukawa couplings from the hierarchical structure of matter field K\"ahler metric \cite{Ishiguro:2021drk}. 
Such a hierarchical structure arises from the coexistence of both the positive and negative modular weights for matter fields. 
Indeed, if the matter fields share the same modular weights, the matter field K\"ahler metric takes a moduli-independent form whose 
value is of ${\cal O}(1)$. See, Ref.~\cite{Kobayashi:2023qzt}, for a theoretical study about a lepton model building by utilizing the positive and negative modular weights.

Let us study the flavor structure of matter fields.
For simplicity, suppose that the Higgs field belongs to ${\cal A}^{\hat{3}}$, and three generations of 
quarks and leptons are originated from the elements of $\{{\cal A}^{\hat{1}}, {\cal A}^{\hat{2}}, {\cal A}^{\hat{3}} \}$. 
In this case, the eigenvalues of physical Yukawa couplings $Y_{\hat{3}\hat{a}\hat{b}}$ are found as
\begin{align}
{\rm Eigen}(Y_{\hat{3}\hat{a}\hat{b}})= 2{\rm diag}\left( 
-\left(\frac{{\rm Im}(t_3)}{{\rm Im}(t_1)}\right)^2,\,\,
(3+E_4(t_1))\left(\frac{{\rm Im}(t_3)}{{\rm Im}(t_1)}\right)^4,\,\,
\frac{-1+E_4(t_1)}{3+E_4(t_1)}\left(\frac{{\rm Im}(t_3)}{{\rm Im}(t_1)}\right)^2
\right),
\end{align}
up to the contribution from the complex structure moduli. 
By using the explicit form of the Eisenstein function, the ratios of these eigenvalues $r_1$ and $r_2$ are  
evaluated as
\begin{align}
r_1 &\simeq -\frac{1}{4}\left(\frac{{\rm Im}(t_1)}{{\rm Im}(t_3)}\right)^2,
\nonumber\\
r_2 &\simeq
15\left(\frac{{\rm Im}(t_1)}{{\rm Im}(t_3)}\right)^2 \sum_{k=0}^\infty \frac{k^3 q^k}{1-q^k}.
\end{align}
Thus, we can obtain the hierarchical physical Yukawa couplings due to the instanton effects as well as the hierarchical moduli values ${\rm Im}(t_1) \ll {\rm Im}(t_3)$. In this region, the determinant of physical Yukawa couplings is also proportional to $q$-expansion $\sum_{k=0}^\infty \frac{k^3 q^k}{1-q^k}$. 
Since the axion ${\rm Re}(t_1)$ controls the size of CP violation, it would be accessible to the small value of Jarlskog invariant, which is beyond the scope of this paper and will be the subject of future investigation.

Just for simple illustration, we have assumed that the Higgs field belongs to ${\cal A}^{\hat{3}}$.
In general, the Higgs fields may correspond to linear combinations of $\{{\cal A}^{\hat{1}}, {\cal A}^{\hat{2}}, {\cal A}^{\hat{3}} \}$, and coefficients of linear combinations 
may be different between the up and down sectors of the Higgs fields.
That may lead to more realistic mass matrices including mixing angles.
At any rate, a direction of light Higgs pair is fixed by mass terms of the Higgs sector, 
which may be generated by vacuum expectation values of singlets and/or non-perturbative effects.
Such a study is beyond our scope, but it would be interesting to study it in 
realistic CY compactifications in future.

\section{Conclusions and discussions}
\label{sec:con}

In this paper, we studied heterotic string theory with standard embedding 
from the viewpoint of the modular symmetry. 
We found that the 4D low-energy effective action is controlled by 
the $SL(2,\mathbb{Z})$ and its subgroups in the asymptotic regions 
of Calabi-Yau moduli space. 
In asymptotic limits, the holomorphic Yukawa couplings are described by 
holomorphic modular forms, as shown in Sec. \ref{sec:modular_forms}. 
To extract a modular form, we focused on a novel structure 
of instanton numbers, e.g., the same instanton number for a particular modulus 
for the $SL(2,\mathbb{Z})$ modular form. 
In the orbifold limit of the moduli space, the moduli are decomposed to 
untwisted moduli and  twisted ones.
Among the full $Sp(2h+2,\mathbb{Z})$ symmetry, only the modular symmetries 
corresponding to the untwisted (bulk) moduli become manifest, although some symmetries such as permutation symmetries of twisted moduli remain among 
the $Sp(2h+2,\mathbb{Z})$ modular symmetry.
The regime of the moduli space, which we studied in this paper, may be another 
limit, where some subgroups of the $Sp(2h+2,\mathbb{Z})$ modular symmetry 
become manifest.

We also evaluated the physical Yukawa couplings of matter fields in a specific 
CY compactification. 
So far, it was pointed out in Refs. \cite{Blesneag:2018ygh,Ishiguro:2021drk} that the matter field K\"ahler metric plays 
an important role in understanding a hierarchical structure of physical Yukawa couplings. 
We found the coexistence of both the positive and negative modular weights for matter fields.
Such a sign difference of the modular weights leads to the hierarchical structure of matter field K\"ahler metric. 
It was also argued in Ref.~\cite{Kobayashi:2023qzt} that the coexistence of both the positive and negative modular weights 
for matter fields provides a new direction for lepton model building. 
Our finding results reproduce these observations. 
Furthermore, we showed that the instanton corrections also give rise to 
a hierarchical structure thanks to the nature of modular forms.\footnote{Such an exponential suppression was also seen in holomorphic Yukawa couplings of twisted modes in heterotic orbifold models \cite{Hamidi:1986vh,Dixon:1986qv,Burwick:1990tu}.}

In this paper, we examined the holomorphic Yukawa couplings in the 4D low-energy effective action, 
and they transform under the modular symmetry.  
However, there are several unsolved problems:

\begin{enumerate}
    \item The origin of $\Gamma_0(N)$ structure

We have seen the non-trivial structure of instanton numbers, determining the remaining modular symmetry in the low-energy 
effective action. However, our analysis was constrained as a limited class of CY threefolds. 
It would be important to perform a comprehensive study about the modular symmetry in the asymptotic region of CY moduli space. 
It will provide a better understanding of $\Gamma_0(N)$ symmetry for underlying CY threefolds.\footnote{Similar subgroups $\Gamma_0(N)$ were found in specific orbifold limits \cite{Bailin:1993wv,Ishiguro:2023wwf}.} 
In the bottom-up approach, moduli-dependent Yukawa couplings was widely used in the so-called 
modular flavor models. 
(See Refs.~\cite{Kobayashi:2023zzc,Ding:2023htn} and references therein.) 
It would also be fascinating to apply our finding $\Gamma_0(N)$ structure to 
the flavor structure of matter fields. We hope to report on this bottom-up direction in the future. 

\item Stringy selection rule

The 4D low-energy effective field theory derived in this paper seems to have additional coupling selection rules.
Some of them can be originated from the original $Sp(2h+2,\mathbb{Z})$ symplectic modular symmetry. 
Also, the doublet structure was found.
It would be important to study more on stringy selection rules.

\item Siegel modular forms

We have studied the regime, where instanton effects of a single modulus are finite, but 
the other vanish.
In this regime, $SL(2,\mathbb{Z})$ or its subgroup becomes manifest and their 
modular forms appear.
CY threefolds have larger  $Sp(2h+2,\mathbb{Z})$ modular symmetry and many moduli.
It would be interesting to study modular symmetries larger than $SL(2,\mathbb{Z})$ including 
multi-moduli and their Siegel modular forms.\footnote{Some Siegel forms were studied 
explicitly in magnetized D-brane models on $T^6$ compactification \cite{Kikuchi:2023dow}. See also for heterotic orbifold models Refs.~\cite{Baur:2020yjl,Nilles:2021glx}.}
We would study it elsewhere.

\end{enumerate}

The modular symmetry controls not only 3-point couplings of matter fields, but 
also higher order couplings \cite{Abe:2009dr,Kobayashi:2021uam}.
Furthermore, the gauge threshold corrections are constrained by the modular symmetry anomalies \cite{Ibanez:1992hc}, although explicit calculations were done in heterotic orbifold models \cite{Dixon:1990pc}.
Moreover, moduli-dependence of the species scale was also constrained by 
the modular symmetry \cite{Cribiori:2023sch,Castellano:2023aum}.
The modular symmetry is relevant to other aspects.
Hence, the modular symmetry is quite important in most of aspects 
of 4D low-energy effective field theory derived from string theory.
Our analysis has opened various interesting directions to study the modular symmetry 
in CY compactifications.

\acknowledgments

This work was supported in part by Kyushu University’s Innovator Fellowship Program (S.N.), JSPS KAKENHI Grant Numbers JP20K14477 (H.O.), JP22J12877 (K.I.), JP23H04512 (H.O) and JP23K03375 (T.K.).

\appendix 

\section{$\Gamma_0(N)$ modular forms}
\label{sec:app}

In this Appendix, we briefly summarize the modular forms of $\Gamma_0(N)$ rather than $SL(2,\mathbb{Z})$.
For more details, see, e.g., Ref. \cite{DHoker:2022dxx}. 
As discussed before, $\Gamma_0(3)$ modular forms can be realized on specific CY threefolds. 
$\Gamma_0(N)$ is a congruence subgroup of level $N$, which is defined as
\begin{align}
    \Gamma_0(N) = 
    \left\{
    \begin{pmatrix}
        a & b\\
        c & d \\
    \end{pmatrix}
    ,\quad
    c\equiv 0\,({\rm mod}\,N)\,
    \right\}.
\end{align}

Before discussing the modular form of $\Gamma_0(N)$, let us remark the modular forms of $SL(2,\mathbb{Z})$. 
It was known that the Eisenstein series $E_k$ for $k\geq 4$ are holomorphic modular forms of weight $k$, 
but there are no holomorphic modular forms of weight 2. The Eisenstein series $E_2$ is a holomorphic function, but not a modular form.\footnote{A non-holomorphic but modular function of weight 2 is given by $E_2^{\ast} (\tau) := E_2(\tau) - \frac{3}{\pi {\rm Im}(\tau)}$ introduced by Siegel.} 
Indeed, the modular transformation of $E_2$ is given by
\begin{align}
    E_2(\gamma \tau) = (c\tau + d)^2 E_2(\tau) + \frac{12}{2\pi i} c (c\tau + d)
\end{align}
with $\gamma$ being the element of $SL(2,\mathbb{Z})$. 
Note that a modular covariant differential operator $D_k$ acting on modular forms of weight $k$ is 
defined as
\begin{align}
    D_k = \frac{1}{2\pi i}\frac{d}{d\tau} - \frac{k}{12}E_2(\tau),
\end{align}
which maps a modular form of weight $k$ to a modular form of weight $k+2$. 
Thus, $E_2(\tau)$ is regarded as a modular connection rather than a modular form, and it can be rewritten as
\begin{align}
    E_2(\tau) = \frac{1}{2\pi i} \partial_\tau \ln \Delta (\tau)
\end{align}
where $\Delta$ is a weight 12 modular (cusp) form of $SL(2,\mathbb{Z})$. 
Specifically, $\Delta$ is the discriminant of Weierstrass equation, which is 
defined by using the Dedekind $\eta$-function:
\begin{align}
    \Delta (\tau) = (2\pi)^{12} \eta(\tau)^{24}.
\end{align}
This modular function $\Delta$ is convenient to define the holomorphic Eisenstein series for $\Gamma_0(N)$. 

Let us consider the following modular transformation of $\Gamma_0(N)$:
\begin{align}
    \tau \rightarrow \tau^\prime = \frac{\tau}{N \tau + 1}
\end{align}
under which $\Delta(n\tau)$ with $n\in \mathbb{N}$ transforms as
\begin{align}
    \Delta(n\tau) \rightarrow \Delta(n\tau^\prime) = \left(-\frac{N\tau +1}{n\tau}\right)^{12} \Delta \left(-\frac{N}{n} - \frac{1}{n\tau}\right).
    \label{eq:Delta_trf1}
\end{align}
Recalling that the modular transformation of $\Delta(\tau)$ is given by
\begin{align}
    \Delta (\tau) \rightarrow \Delta (\gamma \tau) = (c\tau + d)^{12}\Delta(\tau),
\end{align}
Eq. (\ref{eq:Delta_trf1}) is simplified as
\begin{align}
    \Delta(n\tau^\prime) = (N\tau + 1)^{12} \Delta (n\tau),
    \label{eq:Delta_ntau_trf}
\end{align}
for $n|N$. It indicates that $\Delta(n\tau)$ with $n|N$ is regarded as a weight 12 modular form of $\Gamma_0(N)$.

\subsection{Weight 2}
\label{sec:app_weight2}

By using the modular transformation of $\Delta (n\tau)$ \eqref{eq:Delta_ntau_trf}, 
one can construct a modular form of weight 2 under $\Gamma_0(N)$. 
Let us consider the modular transformation of $E_2(n_1\tau)$ with $n_1|N$:
\begin{align}
    E_2(n_1\tau^\prime) 
    &= \frac{1}{(2\pi i)n_1} \partial_{\tau^\prime} \ln \Delta (n_1\tau^\prime)
    \nonumber\\
    &= \frac{1}{(2\pi i)n_1} \frac{\partial \tau}{\partial \tau^\prime} \partial_\tau \left( 12 \ln (N\tau +1) + \ln \Delta (n_1\tau)\right)
    \nonumber\\
    &= (N\tau +1)^2 E_2(n_1\tau) + \frac{N}{n_1} \frac{12}{2\pi i} (N\tau +1).
\end{align}
It indicates that a linear combination of $E_2(\tau)$ and $E_2(n_1\tau)$ transforms as
\begin{align}
    E_2(\tau^\prime) - c_1 E_2(n_1\tau^\prime) = (N\tau +1)^2 \left(  E_2(\tau) - c_1 E_2(n_1\tau)  \right)
    + \frac{12N}{2\pi i} (N\tau +1)\left( 1 - \frac{c_1}{n_1}\right).
\end{align}
Thus, when the constant $c_1$ is chosen as $c_1 = n_1$ with $n_1|N$, 
$E_2(\tau) - n_1 E_2(n_1\tau)$ is a modular form under $\Gamma_0(N)$ of weight 2. 
Furthermore, the modular transformation of $E_2(n_2\tau) - c_2 E_2(n_3\tau)$ 
with $n_{2,3}|N$ and $c_2$ being a constant
\begin{align}
    E_2(n_2\tau^\prime) - c_2 E_2(n_3\tau^\prime) = (N\tau +1)^2 \left(  E_2(n_2\tau) - c_2 E_2(n_3\tau)  \right)
    + \frac{12N}{2\pi i} (N\tau +1)\left( \frac{1}{n_2} - \frac{c_2}{n_3}\right)
\end{align}
indicates that $E_2(n_2\tau) - \frac{n_2}{n_3}E_2(n_3\tau)$ is a modular form under $\Gamma_0(N)$ of weight 2. 

For instance, there are two linearly independent modular forms of weight 2 under $\Gamma_0(4)$:
\begin{align}
    &E_2(\tau) - 2E_2(2\tau),
    \nonumber\\
    &E_2(2\tau) - 2 E_2(4\tau).
\end{align}

\subsection{Weight 4}
\label{sec:app_weight4}

The Eisenstein series $E_4(\tau)$ is a modular form of weight 4 under $SL(2,\mathbb{Z})$. 
Since $E_4(\tau)$ is written by $E_2(\tau)$
\begin{align}
    E_4(\tau) = -\frac{12}{2\pi i} \frac{d}{d\tau} E_2(\tau) +E_2(\tau)^2,
\end{align}
the modular transformation of $E_4(n\tau)$ under $\Gamma_0(N)$ is obtained by using that of $E_2(n\tau)$. 
It results in
\begin{align}
    E_4(n\tau^\prime) = (N\tau +1)^4E_4(n\tau),
\end{align}
for $n|N$. Thus, $E_4(n\tau)$ is a modular form of weight 4 under $\Gamma_0(N)$ for $n|N$.

\subsection{Weight 6}
\label{sec:app_weight6}

The Eisenstein series $E_6(\tau)$ is a modular form of weight 6 under $SL(2,\mathbb{Z})$. 
Since $E_6(\tau)$ is written by $E_4(\tau)$
\begin{align}
    E_6(\tau) = -3D_4E_4(\tau)=-\frac{3}{2\pi i} \frac{d}{d\tau} E_4(\tau) +E_2(\tau)E_4(\tau),
\end{align}
the modular transformation of $E_6(n\tau)$ under $\Gamma_0(N)$ is obtained by using that of $E_2(n\tau)$ and $E_4(n\tau)$. 
It results in
\begin{align}
    E_6(n\tau^\prime) = (N\tau +1)^6 E_6(n\tau),
\end{align}
for $n|N$. Thus, $E_6(n\tau)$ is a modular form of weight 6 under $\Gamma_0(N)$ for $n|N$. 
In a similar way, higher modular forms can be constructed by acting the modular covariant differential operator 
on modular forms of lower modular forms. 

\bibliography{referencesv2}{}
\bibliographystyle{JHEP}

\end{document}